\documentclass[12pt]{iopart}
\pdfoutput=1 
\usepackage{iopams}
\usepackage{graphicx}
\usepackage[colorlinks=true,
            urlcolor=blue,
            linkcolor=blue,
            citecolor=blue]{hyperref}
\usepackage[T1]{fontenc}
\begin{document}
\title[Greybody factors for higher-dimensional non-commutative black
  holes]{Greybody factors for higher-dimensional non-commutative 
geometry inspired black holes} 

\author{Zachary Cox and Douglas M. Gingrich\footnote{Also at TRIUMF,
Vancouver, BC V6T 2A3 Canada} 
}
\address{Department of Physics, University of Alberta, Edmonton, AB T6G
2E1 Canada}
\eads{\mailto{zcox@ualberta.ca}, \mailto{gingrich@ualberta.ca}}

\vspace{10pt}
\begin{indented}
\item[]\today
\end{indented}

\begin{abstract}
Greybody factors are computed for massless fields of spin 0, 1/2,
1, and 2 emitted from higher-dimensional non-commutative geometry
inspired black holes. 
Short-range potentials are used with path-ordered matrix exponentials
to numerically calculate transmission coefficients.
The resulting absorption cross sections and emission spectra are
computed on the brane and compared with the higher-dimensional
Schwarzschild-Tangherlini black hole.
A non-commutative black hole at its maximum temperature in seven
extra dimensions will radiate a particle flux and power of 0.72-0.81
and 0.75-0.81, respectively, times lower than a
Schwarzschild-Tangherlini black hole of the same temperature. 
A non-commutative black hole at its maximum temperature in seven extra
dimensions will radiate a particle flux and power of 0.64-0.72 and
0.60-0.64, respectively, times lower than a Schwarzschild-Tangherlini
black hole of the same mass. 
\end{abstract}

\vspace{2pc}                                                                      

%%%%%%%%%%%%%%%%%%%%%%%%%%%%%%%%%%%%%%%%%%%%%%%%%%%%%%%%%%%%%%%%%%%%%%%
\section{Introduction}

Non-commutative (NC) space-time geometry has allowed insight into the
quantum nature of gravity.
Within the effective theory for NC black holes, the
point-like sources in the energy-momentum tensor, that are normally
represented by Dirac delta functions of position, are replaced by
Gaussian smeared matter distributions of width $\sqrt{2\theta}$. 
Effective theories for non-commutativity have enable calculations of
black hole properties distinctly different from those of classical
gravity.
The NC black hole has a finite maximum temperature.
A minimum mass and horizon radius exist at which the temperature is
zero and the heat capacity vanishes which may terminate Hawking 
evaporation.
These properties are in clear contradistinction to the classical black
hole with temperature becoming infinite as it approaches zero mass and
horizon radius.

Many studies have elucidated the nature of NC geometry inspired black
holes but little attention has been devoted to the calculation of
greybody factors and their subsequent use in studying absorption cross
sections, and particle and energy spectra.
Models for NC geometry inspired chargeless, non-rotating
black holes were developed by Nicolini, Smailagic and
Spallucci~\cite{Nicolini:2005vd}.
The model was extended to the case of charge in four
dimensions~\cite{Ansoldi:2006vg}, generalized to higher dimensions by 
Rizzo~\cite{Rizzo:2006zb}, and then to charge in higher 
dimensions~\cite{Spallucci:2008ez}.
A review of the developments can be found in
Ref.~\cite{Nicolini:2008aj}. 
The Hawking effect and other thermodynamic aspects of NC black holes
have been studies in Ref.~\cite{Banerjee:2008gc,Nozari:2008rc,Banerjee:2009xx}.
Phenomenological considerations for searches with the Large Hadron
Collider (LHC) experiments appeared in Ref.~\cite{Gingrich:2010ed}.
Graybody factor calculations for massless scalar emission were
presented in Ref.~\cite{Nicolini:2011nz}.

The aim of this paper is to present the emission spectra from
non-rotating NC inspired black holes in higher dimensions for massless
fields of spin 0, 1/2, 1, and 2.
The results are compared with the higher-dimensional
Schwarzschild-Tangherlini (ST) black hole.
One of our goals is to separate the temperature characteristics from
the transmission factors in the comparison of emission by performing
calculations of the graybody factors for different black hole masses.

Throughout, we will work in units of $\hbar = c = k = 1$, and use the
PDG~\cite{PDG} definition for the higher-dimensional ADD Planck scale
$M_D$.
Note that $M_*^{n+2} = 8\pi/(2\pi)^nM_D^{n+2}$ is often used in the
literature. 
When $M_*\sim 1$~TeV is taken, the value of $M_D$ will be more than four
times beyond the current experimental lower bounds on $M_D$ for $n \ge
3$, and ruled out for $n = 1$ and 2 extra dimensions.
To allow comparisons with the literature, we have taken the common
values of $M_D\sim~ \sqrt{\theta}^{-1} \sim 1$.
For $n = 0$, the units are $M_D\sim 10^{16}$~TeV and
$\sqrt{\theta}\sim 10^{-35}$~m, and for $n > 0$, and units can be
chosen as $M_D\sim 1$~TeV and $\sqrt{\theta}\sim 10^{-4}$~fm.

%%%%%%%%%%%%%%%%%%%%%%%%%%%%%%%%%%%%%%%%%%%%%%%%%%%%%%%%%%%%%%%%%%%%%%%
\section{Non-commutative geometry inspired black holes}

A nice review of NC geometry inspired black holes already
exists~\cite{Nicolini:2008aj}. 
We will only present the mathematical results used here.
The $g_{00}$ component of the NC inspired metric is

\begin{equation}
h(r) = 1 - \frac{1}{k_n} \frac{M}{M_D} \frac{1}{(M_D r)^{n+1}} P\left(
\frac{n+3}{2}, \frac{r^2}{4\theta} \right)\, ,
\end{equation}

\noindent
where

\begin{equation}
k_n = \frac{n+2}{2^n\pi^{(n-3)/2}\Gamma\left(\frac{n+3}{2} \right)}
\end{equation}

\noindent
and $P$ is the normalized lower-incomplete gamma function

\begin{equation}
P\left( \frac{n+3}{2}, \frac{r^2}{4\theta} \right)  =  
\frac{1}{\Gamma\left( \frac{n+3}{2} \right)} \gamma\left( \frac{n+3}{2},
\frac{r^2}{4\theta} \right)
= \frac{1}{\Gamma\left( \frac{n+3}{2} \right)}
\int_0^\frac{r^2}{4\theta} \mathrm{d}t e^{-t} t^{\frac{n+3}{2}-1}\, .
\end{equation}

\noindent
The symbols are $M$ for the mass of the black hole and $r$ for the radial
distance from the center of the black hole.

For the effective potentials to be discussed soon, we will also
require the derivative of the metric function with respect to $r$: 

\begin{equation}
\fl
h(r)^\prime = \frac{M}{k_n \Gamma\left(\frac{n+3}{2}\right)}
\frac{1}{(M_D r)^{n+2}} \left[ (n+1) \gamma\left( \frac{n+3}{2},
  \frac{r^2}{4\theta}  \right) 
- 2 \left(
  \frac{r^2}{4\theta} \right)^\frac{n+3}{2}  e^\frac{-r^2}{4\theta}
  \right]\, . 
\end{equation}

\noindent
These expressions reduce to the higher-dimensional commutative forms in
the limit 
$\theta\to 0$ or $P\to 1$. 

The horizon radius for non-commutative inspired black holes is given by

\begin{equation}
\frac{M}{M_D} = \frac{k_n}{P \left( \frac{n+3}{2},
\frac{r_\mathrm{h}^2}{4\theta} \right)} (M_D r_\mathrm{h})^{n+1}\, .  
\end{equation}

\noindent
We are unaware of a closed form solution to this equation and so have
solved it numerically for $r_\mathrm{h}$.
In the commutative limit, the horizon reduces to the usual
higher-dimensional ST case, which can be written as  

\begin{equation}
r_\mathrm{S} = \frac{1}{M_D} \left( \frac{1}{k_n} \frac{M}{M_D}
\right)^\frac{1}{n+1}\, . 
 \end{equation}

\noindent
Depending on the context, $r_\mathrm{h}$ will be used to refer to both
NC and ST black hole horizon radii.

\begin{figure}[htb]
\centering
\includegraphics[width=0.6\textwidth]{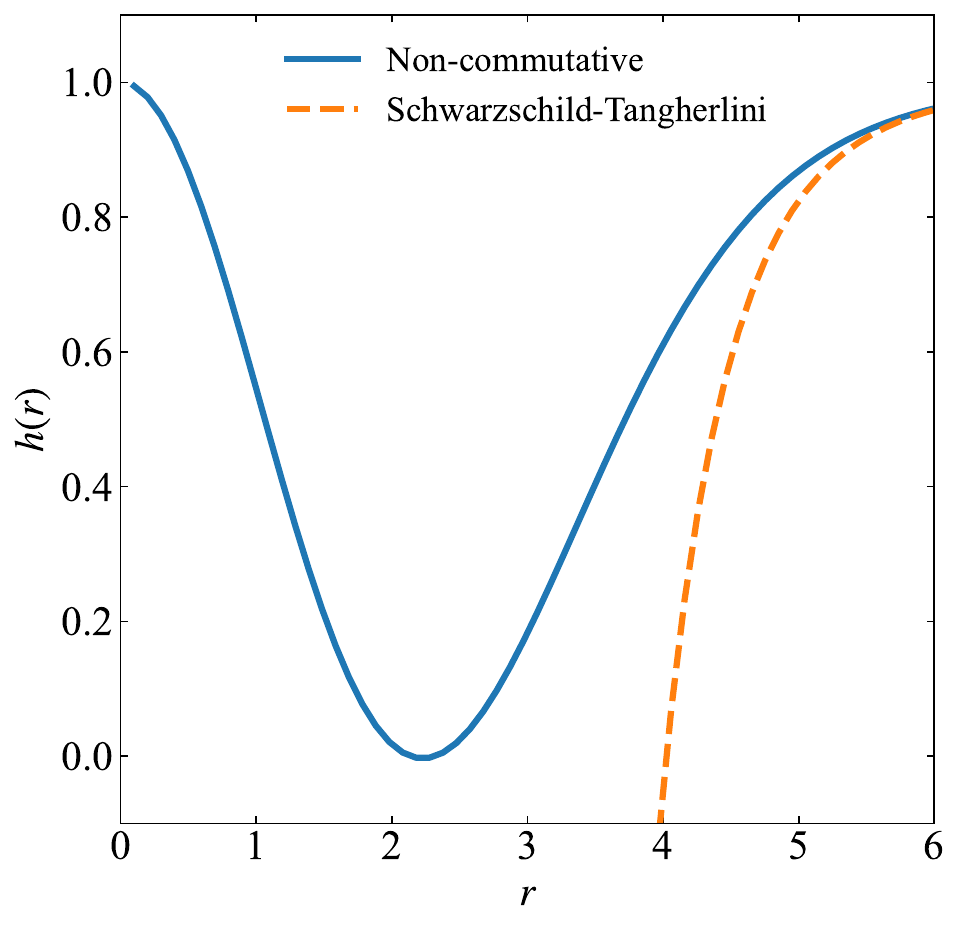}
\caption{\label{fig:horizon}%
Metric function $h(r)$ for non-commutative (solid line) and
commutative (dashed line) black holes with $n=7$, $M = 20.4 M_D$, and
$M_D = \sqrt{\theta} = 1$.
}
\end{figure}

For given $M$, $n$, $M_D$, and $\theta$, there can be one, two, or no
horizon.
For reasons related to positive temperature given later, only the
outer horizon radius is relevant to the work presented here.
Figure~\ref{fig:horizon} shows $h(r)$ for the case of a single event
horizon for $n=7$, $M = 20.4 M_D$, and $M_D = \sqrt{\theta} = 1$, and a
comparison with the ST black hole. 

Because there is a minimum mass, there are masses below which the
black hole will not form, and above the minimum mass the horizon
radius is double valued.
As the mass increases, the inner horizon radius shrinks to zero, while
the outer horizon radius approaches the commutative value.
The situation thus depicted in Figure~\ref{fig:horizon} represents the
minimum mass case in which the values of the horizon radii for NC and
ST black holes are maximally different.
This will be a useful condition when considering maximum differences
in the transmission coefficients between NC and ST black holes.

%%%%%%%%%%%%%%%%%%%%%%%%%%%%%%%%%%%%%%%%%%%%%%%%%%%%%%%%%%%%%%%%%%%%%%%
\section{Transmission coefficients}

Hawking radiation from black holes is typically studied by examining
the response to perturbations. 
Hence, understanding modifications of Hawking radiation due to NC
geometry requires one to solve the equations of motion for various
spin perturbations on the NC inspired black hole metric. 

The Teukolsky equation describes spin 0, 1/2, 1, and 2 field
perturbations in the background metric due to the black
hole~\cite{Teukolsky:1973ha, teukolsky1973part2, Teukolsky:1974yv}.
The partial differential equations can be separated.
The angular equation can be numerically solved to obtain the energy
eigenvalues or separation constant.
The energy eigenvalues are then used in the radial equation.
The radial equation can be solve to find transmission
coefficients for fields emitted from the black hole horizon.
The transmission coefficients describe the probability that a
particle, generated by quantum fluctuations at the horizon of a black
hole, escapes to spatial infinity.

The Teukolsky radial equation can be numerically solved directly, but
the convergence of the solution at the integration boundaries is not
clear.
The difficulty with the radial equation in the context of a scattering
problem is that the first order radial derivatives create complex
$is\omega$ terms which have a $1/r$ behaviour at infinity.
This can eventually lead to problems in the numerical computations.

Another approach uses a Chandrasekhar
transformation~\cite{Chandrasekhar} to cast the
radial equation into an effective Schr{\"o}dinger equation with a
short-range barrier potential different for each spin field, allowing
a more careful numerical treatment. 
The equation takes the form

\begin{equation}
\left[ \frac{d^2}{dr^2_*} + \omega^2 -V_s(r) \right] \psi_s = 0\, ,
\label{eq:schrodinger}
\end{equation}

\noindent
where $s$ is the spin of the field and $r_*$ a generalized tortoise
coordinate.
We are now faced with a potential-barrier problem.

The purpose of this transformation is that the potentials $V_s(r)$ are
now short-ranged.
They vanish faster than $1/r$ which is advantageous for numerical
computations. 
It should be noted that these potentials contain a dependence on
$\omega$ through the connection coefficient to the angular equation.
Working with real-valued potentials has benefits.

Arbey et al.~\cite{Arbey:2021jif} have derived general
potentials $V_s$ for spherical symmetric metrics for different spins $s$
of massless fields.
The effective potentials seen by $s = 0, 1/2, 1$, and 2 massless fields can be
written as 

\numparts
\begin{eqnarray}
V_0     & = & h \left[ \frac{\nu_0}{r^2} + \frac{h^\prime}{r} \right]\, ,\\
V_{1/2} & = & \nu_{1/2} \frac{h}{r^2} \pm \sqrt{\nu_{1/2}h}\left[
    \frac{h^\prime}{2r^2} - \frac{h}{r^2} \right]\, ,\label{eq:potential}\\ 
V_1     & = & \nu_1 \frac{h}{r^2}\, ,\\
V_2     & = & h \left[ \frac{\nu_2+2h}{r^2} - \frac{h^\prime}{r}
\right]
\label{eq:V2}\, ,
\end{eqnarray}
\endnumparts

\noindent
where $\nu_0 = \nu_1 = \ell(\ell+1)$, $\nu_{1/2} = \ell(\ell+1)+1/4$,
$\nu_2 = \ell(\ell+1)-2$, and $\ell$ is the angular momentum
quantum number.

For spin 1/2, one must take the positive sign in Eq.~(\ref{eq:potential}) to
get a positive potential for all $r$.
For the spin-2 case, for high $n$ and $\ell = 2$, it is possible
to get a negative potential.
We will discuss this further below when describing absorption cross
section results.
These potentials differ from those in Ref.~\cite{Arbey:2021jif} in
that they are missing a $1/\sqrt{2}$ factor in the second term of
Eq.~(5.1d), and Eq.~(5.9c) should have a first constant of 4 rather
than 2 in the second term and a constant of 3 rather than 1 in the
last term. 

\begin{figure*}[htb]
\includegraphics[width=\textwidth]{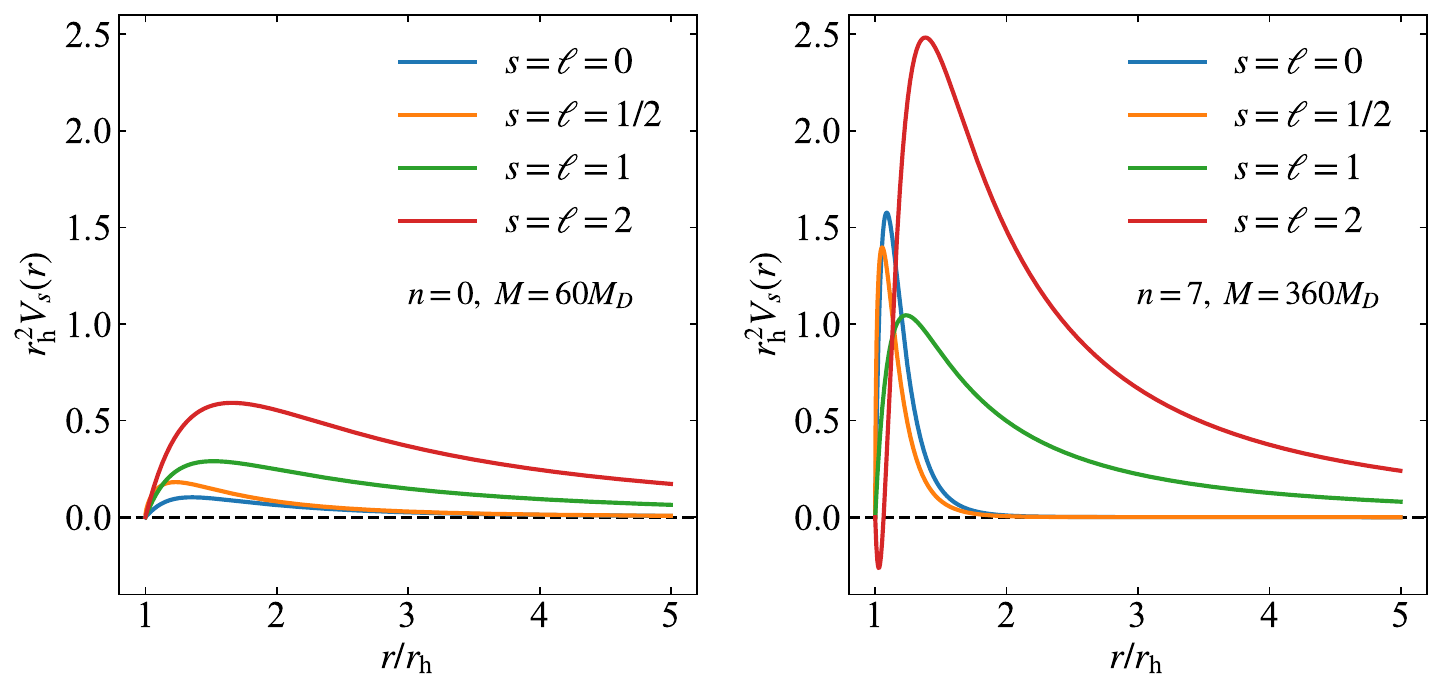}
\caption{\label{fig:potentials}%
Effective potentials for non-commutative black holes for spin
0, 1/2, 1, and 2 massless fields with $M_D = \sqrt{\theta} = 1$:
(left) $n = 0, M = 60 M_D$ and (right) $n = 7, M = 360 M_D$.
}
\end{figure*}

Figure~\ref{fig:potentials} shows two representative cases for the NC
black hole potentials with $M_D = \sqrt{\theta} = 1$ for spin 0,
1/2, 1, and 2; the case of $l = s$ is shown.

The Schr{\"o}dinger Eq.~(\ref{eq:schrodinger}) is in the tortoise
coordinate $r_*$ while the potential is in $r$; or one can view $r =
r(r_*)$. 
The relationship between the coordinates  is defined by

\begin{equation}
\frac{\mathrm{d}r_*}{\mathrm{d}r} = \frac{1}{h(r)}\label{eq:tortoise}\, .
\end{equation}

\noindent
We note that at $r\to+\infty$, $r_*\to+\infty$, and as $r\to
r_\mathrm{h}$, $r_*\to-\infty$.

One thus needs to solve differential Eq.~(\ref{eq:tortoise}) for $r_*$
and also invert it to obtain $r$.
While the result is well known for the four-dimensional Schwarzschild
metric it is less apparent for others.
Analytic expressions exist for $r_*$ for the higher dimensional
ST case but we are unaware of such analytic
expressions for the NC case.
We have thus numerically integrated Eq.~(\ref{eq:tortoise}).
For the initial condition, we take $r=r_*$ for a very large value
(approximating $+\infty$), and integrate backwards.
The procedure has been validated in the ST case by comparing the
numerical integrations with the analytic results.  
The analytic formulae are taken from Ref.~\cite{Arbey:2021yke}, to
which we have added the $n=7$ case:

\begin{eqnarray}
\fl
r_* = r + \frac{r_\mathrm{h}}{16} \left[ \sqrt{2} \ln \frac{x^2 -
    \sqrt{2}x + 1}{x^2 + \sqrt{2}x + 1} + 2\ln \frac{x-1}{x+1}
  + 4\tan^{-1}\frac{1}{x}\right.\nonumber\\
\left.- 2\sqrt{2} \tan^{-1} \frac{1}{\sqrt{2}x-1}
+ 2\sqrt{2} \tan^{-1} \frac{1}{\sqrt{2}x+1}
\right]\, ,   
\end{eqnarray}

\noindent
where $x = r/r_\mathrm{h}$.
We note that Eq.~(A5) in Ref.~\cite{Arbey:2021yke} is missing an
essential negative sign under the square-root for the first
$\sqrt{\varphi_-}$. 

We integrate the tortoise equation using a variable step size over the
range $r/r_\mathrm{h} = [\epsilon,350]$, where $\epsilon \approx 10^{-16}$.
To obtain the inverse relation for $r$ in terms of $r_*$, we have
integrated the tortoise equation and numerically inverted it using
linear interpolation. 

Gray and Visser~\cite{Gray:2015xig} showed that the Bogoliubov
coefficients relating incoming and outgoing waves on a potential
barrier can be directly obtained from the following path-ordered exponential 

\begin{equation}
\left[
\begin{array}{cc}
\alpha & \beta^*\\
\beta & \alpha^*
\end{array}
\right]
= \mathcal{P} \exp \left(
-\frac{i}{2\omega} \int_{-\infty}^{+\infty}
V_s(r_*)
\left[
\begin{array}{cc}
1 & e^{-2i\omega r_*}\\
 -e^{2i\omega r_*} & -1
\end{array} \right] dr_*\right)\, ,
\end{equation} 

\noindent
where $\mathcal{P}$ is a path-ordering operator.
Using the product calculus definition of path-ordered integrals, they
compute the Bogoliubov coefficients via the product integrals

\begin{equation}
\left[
\begin{array}{cc}
\alpha & \beta^*\\
\beta & \alpha^*
\end{array}
\right]
= \prod_{+\infty}^{-\infty} \left[ I + A(r_*)dr_*\right]\, ,
\end{equation}

\noindent
where $I$ is the identity matrix and $A(r_*)$ is the transfer
matrix given by

\begin{equation}
A(r_*) = -\frac{i}{2\omega} V_s(r_*) \left[ \begin{array}{cc} 1 &
    e^{-2i\omega r_*}\\ -e^{2i\omega r_*} & -1\end{array}\right]\, .
\end{equation}

The product integral can be approximated numerically by

\begin{equation}
\left[
\begin{array}{cc}
\alpha & \beta^*\\
\beta & \alpha^*
\end{array}
\right]
= \lim_{N\to\infty} \left[ (I+A((r_*)_{N-1})h)\ldots 
(I+A((r_*)_1)h)\right]\, ,
\label{eq:integral}
\end{equation}

\noindent
where $(r_*)_i > (r_*)_{i-1}$ and $h$ is the step size.
We have taken $N = 10^4$.

The transmission probabilities $\Gamma(\omega)$ are related to the
Bogoliubov coefficients by

\begin{equation}
\Gamma(\omega) = \frac{1}{|\alpha(\omega)|^2}\, .
\end{equation}

\noindent
By using this procedure, one does not actually solve numerically
a differential equation. 
The problem becomes one of performing a single numerical integral,
Eq.~(\ref{eq:integral}).

%%%%%%%%%%%%%%%%%%%%%%%%%%%%%%%%%%%%%%%%%%%%%%%%%%%%%%%%%%%%%%%%%%%%%%%%%%%%%%%%%%
\section{Hawking emission}

In studying Hawking emission from NC geometry inspired black holes, we
consider the following.
The absorption cross section is an observable acting as an effective area
representing the likelihood of a particle to be scattered by the black hole:

\begin{equation}
\sigma_s(\omega) = \frac{\pi}{\omega^2} \sum_{\ell\ge s} (2\ell + 1)
\Gamma_{s,l}(\omega)\, ,
\label{eq:cross}
\end{equation}

\noindent
\noindent
where $\Gamma_{s,\ell}(\omega)$ are transmission coefficients,
greybody factors, for angular momentum mode $\ell$, and $2\ell + 1$ is a
multiplicity factor for the azimuthal modes $m$ in spherical geometry.

One of the most interesting aspects of NC inspired black holes is their
temperature properties. 
The temperature in terms of the horizon radius is given by

\begin{equation}
T = \frac{n+1}{4\pi r_\mathrm{h}} \left[ 1 - \frac{2}{n+1} \left(
\frac{r_\mathrm{h}}{2\sqrt{\theta}} \right)^{n+3}
\frac{e^{-r_\mathrm{h}/(4\theta)}}{\gamma \left( \frac{n+3}{2},
\frac{r_\mathrm{h}^2}{4\theta}\right) }  \right]\, . 
\end{equation}

\begin{figure}[htb]
\centering
\includegraphics[width=0.6\textwidth]{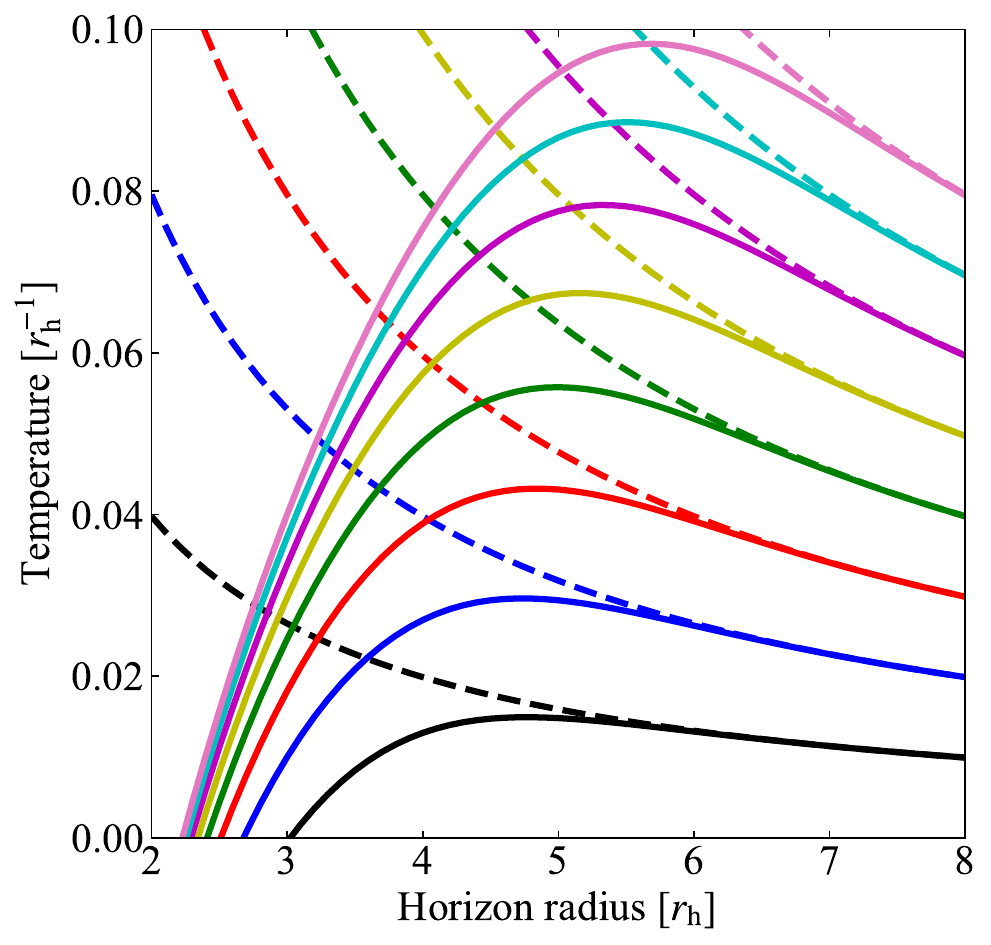}
\caption{\label{fig:temperature}%
Black hole temperature versus horizon radius for different number of
extra dimensions with $M_D = \sqrt{\theta} = 1$.
The solid lines are for non-commutative inspired black holes and
the dashed lines for Schwarzschild-Tangherlini black holes.
The number of extra dimensions increases from 0 (bottom) to 7 (top).
}
\end{figure}

\noindent
The quantity in square brackets modifies the usual higher-dimensional
commutative form.
In addition, since $r_\mathrm{h}$ has been modified, it also leads to
temperature differences.
Since the inner horizon radius corresponds to negative temperature, we
only considered the outer radius.
Figure~\ref{fig:temperature} shows the temperature versus horizon
radius for NC black holes with $M_D = \sqrt{\theta} = 1$.
The temperature vanishes at the minimum mass and there is a maximum
temperature.
Also shown for comparison are the usual temperatures for ST black
holes. 

Of interest to us will be the NC black hole maximum temperature
$T_\mathrm{max}$ and mass $M_{\mathrm{e}M}$ at which the maximum
temperature occurs.  
In addition, we will use values of the ST black hole mass
$M_\mathrm{eT}$ at which the temperature is the same as the NC black 
hole maximum temperature.
These values are shown in Table~\ref{tab:temperature}. 

\begin{table}[htb]
\caption{\label{tab:temperature}%
Non-commutative inspired black hole maximum temperature
$T_\mathrm{max}$, mass at which the maximum temperature occurs
$M_{\mathrm{e}M}$, and mass $M_\mathrm{{e}T}$ of a
Schwarzschild-Tangherlini black hole that has temperature
$T_\mathrm{max}$, for different number of extra dimensions $n$ with
$M_D = \sqrt{\theta} = 1$.
}
\begin{indented}
\item[]\begin{tabular}{ccccccccc}
\br
$n$ & 0 & 1 & 2 & 3 & 4 & 5 & 6 & 7\\
\mr
$T_\mathrm{max}$ & 0.015 & 0.030 & 0.043 & 0.056 & 0.067 & 0.078 & 0.089 &
  0.098\\
$M_{\mathrm{e}M}$ & 60.47 & 108.3 & 157.5 & 204.9 & 249.1 & 289.4 & 325.6 &
  357.7\\
$M_{\mathrm{e}T}$ & 66.95 & 136.0 & 224.8 & 332.0 & 456.4 & 597.0 & 753.0 &
  923.7\\
\br
\end{tabular}
\end{indented}
\end{table}

We now introduce the main physical variables that can be formulated
using the transmission coefficients.
The number of particles emitted per unit time and per unit frequency
is

\begin{equation}
\frac{\mathrm{d}^2N}{\mathrm{d}t \mathrm{d}\omega} = \frac{1}{2\pi}
\frac{1}{\exp(\omega/T) - (-1)^{2s}} \sum_{\ell\ge s}
(2\ell+1)\Gamma_{s,\ell}(\omega)\, . 
\label{eq:particle}
\end{equation}

\noindent
The energy emitted per unit time (or power) and per unit frequency is
\begin{equation}
\frac{\mathrm{d}^2E}{\mathrm{d}t \mathrm{d}\omega} = \frac{1}{2\pi}
\frac{\omega}{\exp(\omega/T) - (-1)^{2s}} \sum_{\ell\ge s}
(2\ell+1)\Gamma_{s,\ell}(\omega)\, . 
\label{eq:energy}
\end{equation}

\noindent
We acknowledge a damping factor $\exp(-\theta\omega^2/2)$ that was
developed in Ref.\cite{Nicolini:2011nz} that multiplies
Eq.~(\ref{eq:particle}) and Eq.~(\ref{eq:energy}).
Including this factor gives a sub-dominate
effect~\cite{Nicolini:2011nz} which is not particularly relevant to
our discussion and will be ignored.

The black body radiation leaving the horizon sees an effective 
potential barrier due to the geometry of the space-time surrounding
the black hole.
The potential barrier attenuates the radiation such that an observer at
spatial infinity away from the black hole will measure a different
emission spectrum than the one at the horizon by a factor
$\Gamma_{s,l}(\omega)$ called the greybody factor.
Thus the grey body factor $\Gamma_{s,l}(\omega)$ represents the
fraction of the black body emission which penetrates through the
potential barrier and escapes to spatial infinity.  
The expressions for absorption cross section Eq.~(\ref{eq:cross}) and
emission rates Eq.~(\ref{eq:particle}) and Eq.~(\ref{eq:energy}) are
only applicable to radiation on the brane.  

%%%%%%%%%%%%%%%%%%%%%%%%%%%%%%%%%%%%%%%%%%%%%%%%%%%%%%%%%%%%%%%%%%%%%%%%%%%%%%%%%%
\section{Results}

We present calculations of transmission coefficients, absorption cross
sections, and spectra for spin 0, 1/2, 1, and 2 fields.
We have validated the method by comparing with well known results for
four dimensional Schwarzschild and Kerr black holes, as well as the ST
black holes~\cite{Harris:2003eg,Park:2005vw} that we use for
comparison with the NC results. 

The model of NC geometry inspired black holes in higher dimensions has
three unknown parameters $n$, $M_D$, and $\sqrt{\theta}$.
Typically we present results for each extra dimension $n$.
Usually it is necessary to fix the other two parameters.
One possibility for fixing the parameters is to be guided by
experimental constraints.
Updating the approach taken in Ref.~\cite{Gingrich:2010ed} (see the
Appendix) restricts the values of $\sqrt{\theta}M_D$ that
%Appendix\ref{sec:LHC}) restricts the values of $\sqrt{\theta}M_D$ that
can be probed from 0.25 to 0.98, different for each number of extra
dimensions.
The allowed range in $\sqrt{\theta}M_D$ for any given number of extra
dimensions is severely restricted. 

Calculations using values of $\sqrt{\theta}M_D< 1$ begin to probe the
details of the matter smearing distribution and become model
dependent. 
However, the primary goal in this paper is to study the differences in
Hawking emission from NC and ST black holes so we choose the usual
condition $M_D = \sqrt{\theta} = 1$.
This implies that our phenomenological predictions will not have
particular consequence for the physics at the LHC. 

%%%%%%%%%%%%%%%%%%%%%%%%%%%%%%%%%%%%%%%%%%%%%%%%%%%%%%%%%%%%%%%%%%%%%%%%%%%%%%%%%%
\subsection{Transmission coefficients}

\begin{figure*}[htb]
\includegraphics[width=\linewidth]{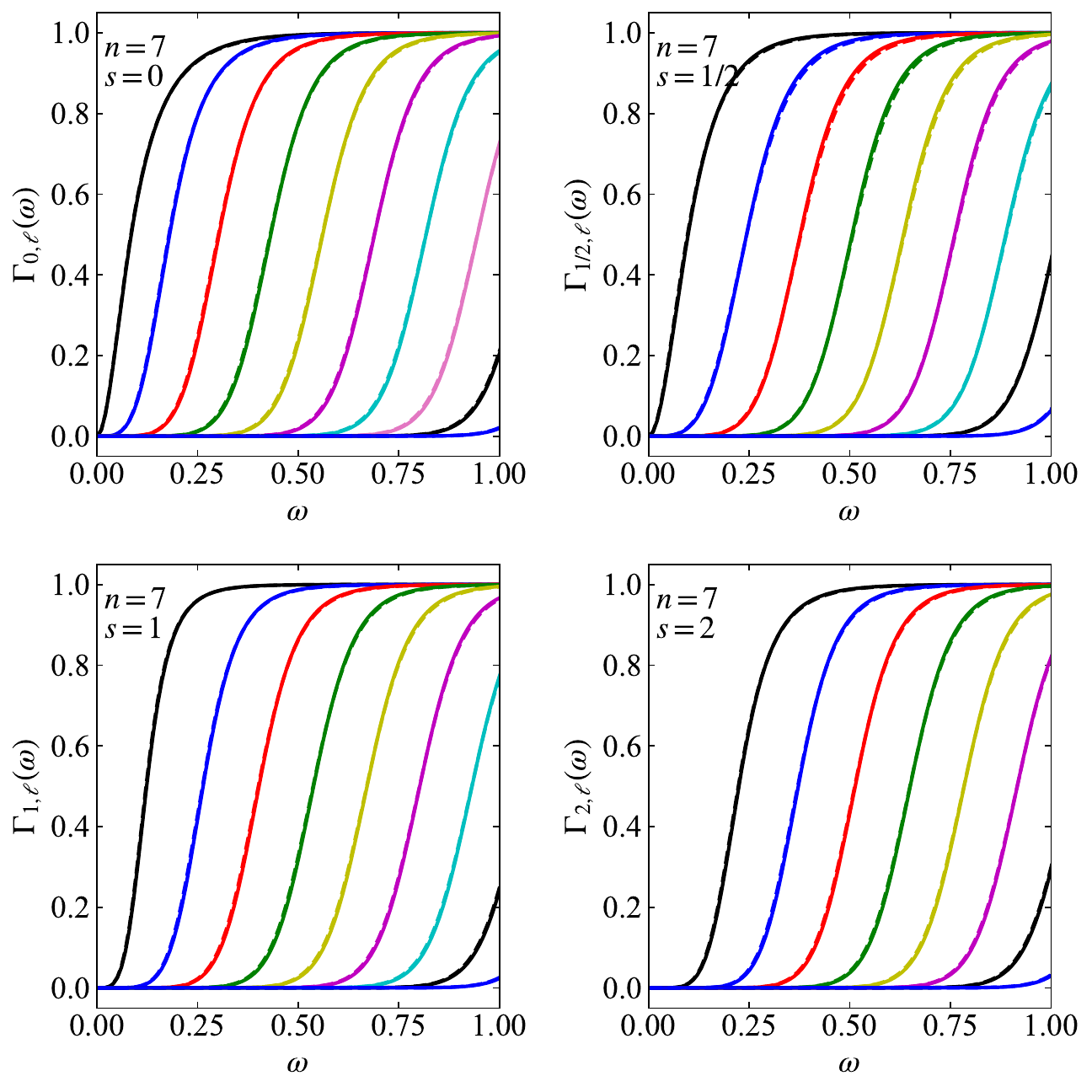}
\caption{\label{fig:transeM}%
Transmission coefficients $\Gamma_{s,\ell}(\omega)$ for $s = 0, 1/2,
1, 2$ as a function of frequency $\omega$.
The quantum number $\ell$ increases from $\ell = s$ going from left to right.
The solid lines are for non-commutative black holes and dashed lines
for Schwarzschild-Tangherlini black holes.
A black hole mass of $358 M_D$ has been used and $M_D = \sqrt{\theta} = 1$ taken.
}
\end{figure*}

The fundamental calculated quantity is the transmission coefficient
as a function of frequency for different black hole masses, number of
extra dimensions, spin, and $\ell$ modes.  
Figure~\ref{fig:transeM} shows transmission coefficients for $n = 7$ and
$s = 0, 1/2, 1, 2$ as a function of frequency for different $\ell$ modes. 
The solid lines are for NC black holes and dashed lines
for ST black holes.
The quantum number $\ell$ increases from $\ell = s$ going from left to right.
A black hole mass of $358 M_D$ is used and corresponds to the NC black hole
maximum temperature.
We observe that the NC and ST black hole transmissions coefficients at
this mass are very similar, differing slightly for higher $l$.
This is because they have a similar horizon radius at this mass of
5.68 for NC black holes and 5.76 for ST black holes.

\begin{figure*}[htb]
\includegraphics[width=\linewidth]{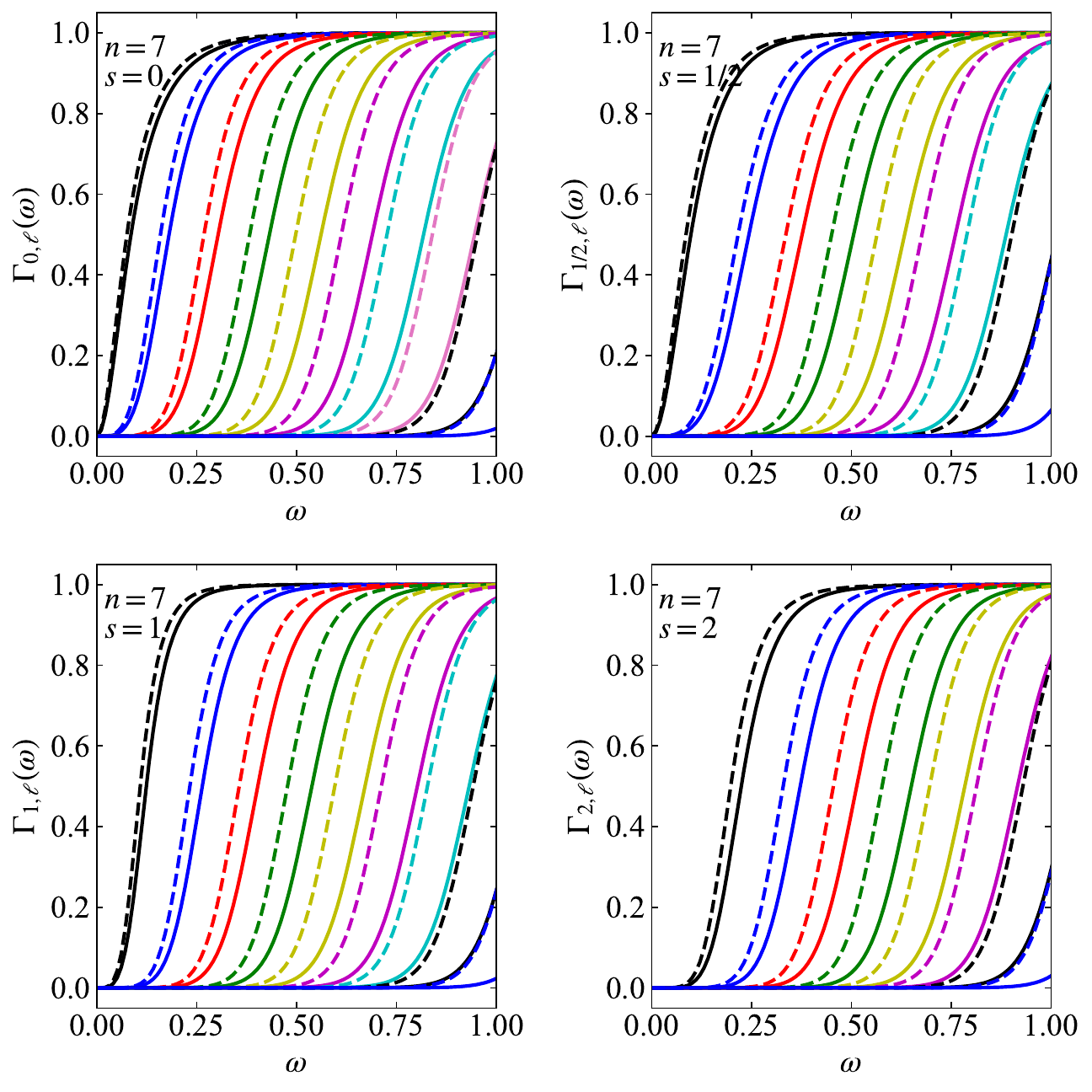}
\caption{\label{fig:transeT}%
Transmission coefficients $\Gamma_{s,\ell}(\omega)$ for $s = 0, 1/2,
1, 2$ as a function of frequency $\omega$.
The quantum number $\ell$ increases from $\ell = s$ going from left to right.
The solid lines are for non-commutative black holes $M = 358 M_D$ and dashed lines
for Schwarzschild-Tangherlini black holes $M = 924 M_D$.
A black hole temperature of $0.098 M_D$ has been used and $M_D
= \sqrt{\theta} = 1$ taken. 
}
\end{figure*}

If the horizon radius difference between NC and ST black holes is
significantly different, the comparison changes.
Figure~\ref{fig:transeT} shows transmission coefficients for $n = 7$ and
$s = 0, 1/2, 1, 2$ as a function of frequency for different $\ell$ modes. 
A black hole mass of $358 M_D$ and $924 M_D$ are used for the NC black
hole and ST black hole, respectively, corresponding to the NC black
hole maximum temperature.
In this case, significant differences are observed for a horizon radius
of 5.68 for NC black holes and 6.48 for ST black holes.

\begin{figure*}[htb]
\includegraphics[width=\linewidth]{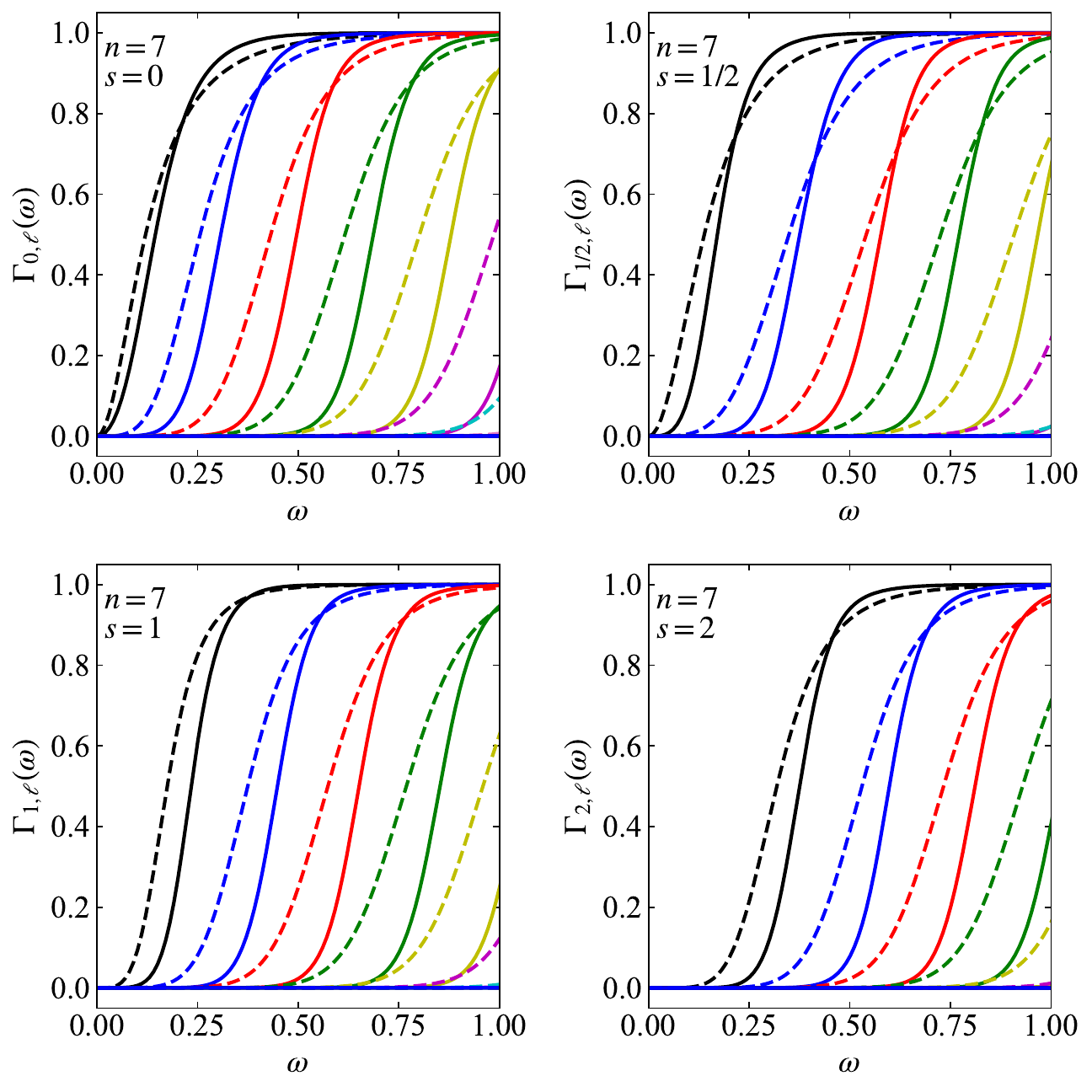}
\caption{\label{fig:trans1}%
Transmission coefficients $\Gamma_{s,\ell}(\omega)$ for $s = 0, 1/2,
1, 2$ as a function of frequency $\omega$.
The quantum number $\ell$ increases from $\ell = s$ going from left to right.
The solid lines are for non-commutative black holes and dashed lines
for Schwarzschild-Tangherlini black holes.
A black hole mass of $20.4 M_D$ has been used and $M_D = \sqrt{\theta} = 1$ taken.
}
\end{figure*}

To examine more significant differences in transmission coefficients,
a value for the black holes mass at the minimum NC black hole mass
can be chosen and is shown in Figure~\ref{fig:trans1}.
The horizon radius of the NC black hole is 2.32 and that of the
ST black hole 4.02.
We observe significant differences between NC and ST black hole
transmission coefficients with increasing $\ell$.
The NC black hole transmission probabilities begin to rise at higher
frequencies but rise more steeply than the ST transmission
probabilities. 
This behaviour was first observed for spin 0 in Ref.~\cite{Nicolini:2011nz}.

The number of effective $\ell$ modes used in the calculations can
vary.  
The total number of $\ell$ modes considered are 15, 15, 14, 13
for $s=0,1/2,1,2$, respectively.
The number of effective $\ell$ modes giving a non-negligible
contribution in the frequency range $0 < \omega \le 1$ is different
depending on $M$, $n$, and $s$.
Typically, $s=0$ and 1/2 have the same number of effective modes,
while $s=1$ has one less and $s=2$ has two less modes.
The $s=1$ and 2 cases have transmission coefficients that turn-on at
higher frequencies relative to the $s=0$ and 1/2 cases, i.e.\ because
of $\ell \ge s$, the higher spins are missing the lower $\ell$ modes.
As $n$ increases, the transmission coefficients become more spread
out, and thus less modes will contribute to the given frequency range.
The $n=0$ case has about four more modes than $n=7$.
The lowest masses we consider will have about three less effective
modes than the highest masses we consider.
The number of effective modes that will fit into the frequency range
is largely determined by the spacing of the transmission coefficients
in frequency.

Another important characteristic of the transmission coefficients is
how steeply they rise with increasing frequency.
In general, the turn-on steepness is largely independent of spin
except for the $\ell=0$ and 1 modes.
The more effective number of modes, the steeper the turn-on.
Visually, the turn-on is most steep for $s=1$ and less step for $s=2$.

The differences in transmission coefficients between NC and ST black
holes depend significantly on their relative horizon radii.
Typically a bigger horizon radius will give transmission coefficients
that turn-on lower in frequency; the difference becoming more
pronounced as $\ell$ increases.
In addition, it is observed that at lower masses although the ST black
hole transmission coefficients turn-on sooner, the NC black hole
coefficients rise steeper and become higher before plateauing to unity,
especially for $s=1/2$. 

%%%%%%%%%%%%%%%%%%%%%%%%%%%%%%%%%%%%%%%%%%%%%%%%%%%%%%%%%%%%%%%%%%%%%%%%%%%%%%%%%%
\subsection{Absorption cross sections}

\begin{figure*}[htb]
\includegraphics[width=\linewidth]{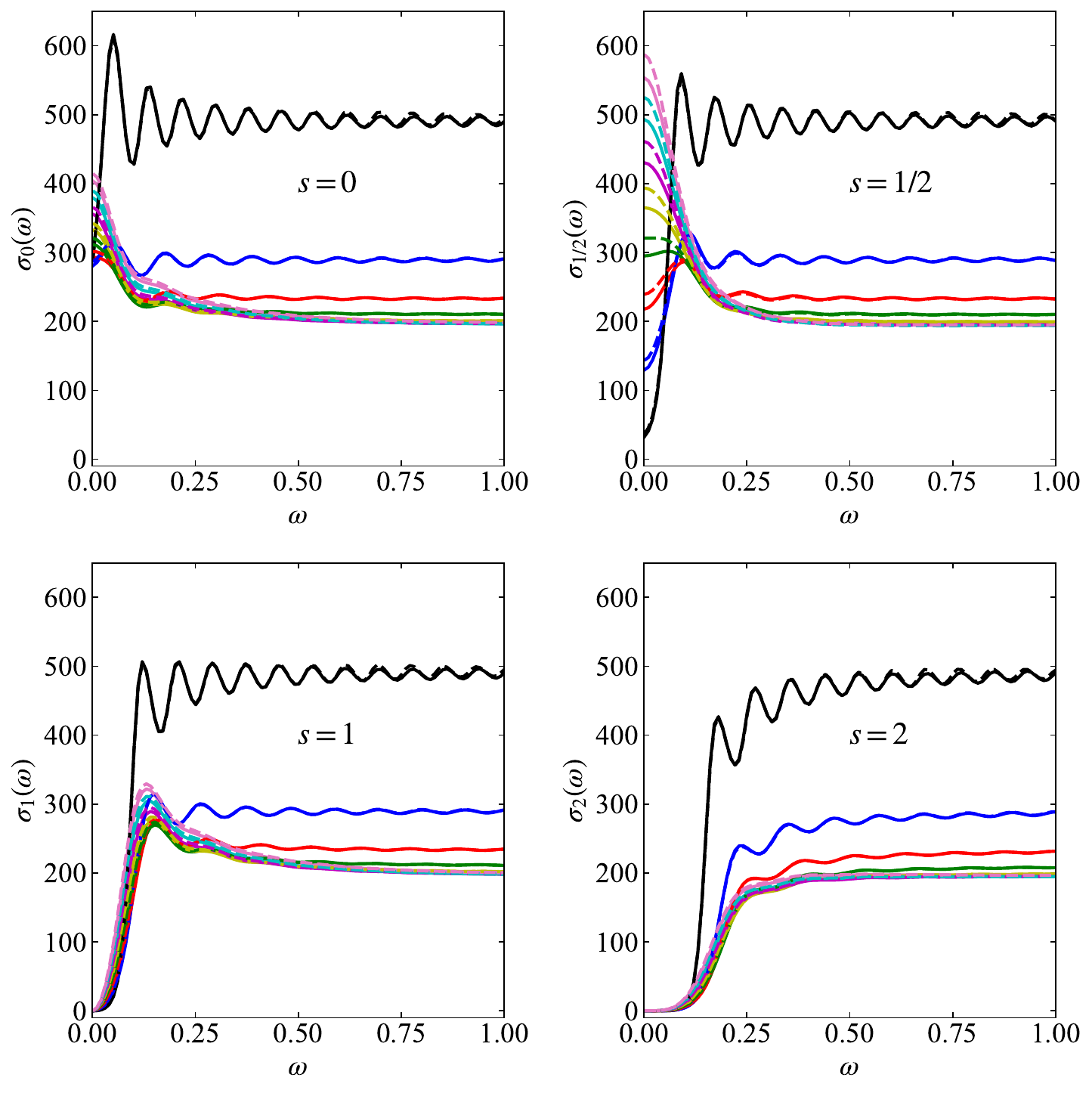}
\caption{\label{fig:cross0}%
Absorption cross sections versus frequency $\omega$ for $s = 0, 1/2, 1, 2$.  
The solid lines are for non-commutative black holes and dashed lines
for Schwarzschild-Tangherlini black holes.
The number of extra dimensions increases from 0 to 7 as the curves
moved from top to bottom at high $\omega$.
Black hole masses corresponding to the non-commutative black hole maximum
temperature have been used.
$M_D = \sqrt{\theta} = 1$ has been taken.
}
\end{figure*}

The absorption cross section depends on the weighted sum over $\ell$ of
transmission coefficients and inversely as $1/\omega^2$.
Figure~\ref{fig:cross0} shows cross sections versus frequency for $s =
0, 1/2, 1, 2$.
The solid lines are for NC black holes and dashed lines for ST black holes.
Black hole masses corresponding to the NC black hole maximum
temperature have been used; equal NC and ST black hole masses,
$M_{\mathrm{e}M}$ in Table~\ref{tab:temperature}. 
Differences in cross sections are observed at low frequencies.
These differences are most significant for $s=1/2$ and less pronounced
for $s=2$.

Hawking radiation for spin-2 fields in the ST metric was first
discussed by Park~\cite{Park:2005vw}.
Direct comparison is not possible since we use a different effective
potential which is taken from Ref.~\cite{Arbey:2021jif} but originally
comes from Ref.~\cite{Moulin:2019ekf}. 
The difference in the general form of the potential appears
significant but when substituting the particular ST metric, the
difference is replacing the $-1$ coefficient of the second term in
Eq.~(\ref{eq:V2}) by $-(n+1)$. 
Noteworthy in Ref.~\cite{Park:2005vw} is the acknowledgement that
the spin-2 potential can become negative -- potential well -- for some 
masses (or radii), number of extra dimensions, and $\ell$ modes. 
The potential well can occur in the region $r_* \sim 0$. 
For the ST metric, the condition for non-negative potential is $n\le 3$.
If $4 \le n \le 7$, the $\ell = 2$ mode feel a potential well.
The depth of the potential well, and height of the barrier, increase
with increasing number of extra dimensions.
A trade-off can occur between barrier suppression and well
enhancement.
For the ST case,  this causes the $n=5$--7 spin-2 cross sections to
rise slightly faster at low-frequency than the $n=0$--4 cross
sections. 
The same observations are made for the NC case.
However, the effect is small and will not concern use for the
remainder of this paper.

\begin{figure*}[htb]
\includegraphics[width=\linewidth]{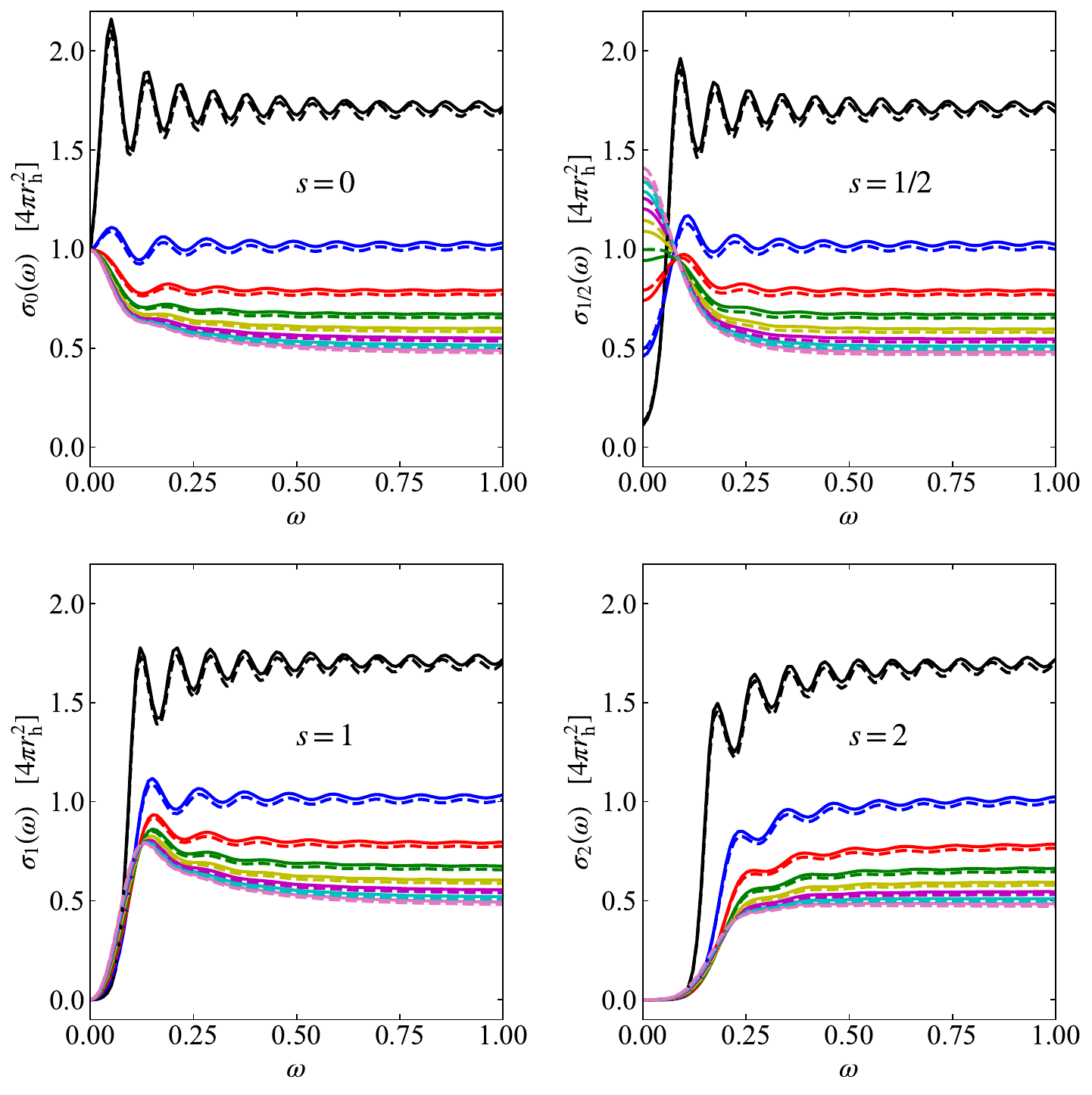}
\caption{\label{fig:cross1}%
Normalized absorption cross sections versus frequency $\omega$ for $s
= 0, 1/2, 1, 2$.   
The solid lines are for non-commutative black holes and dashed lines
for Schwarzschild-Tangherlini black holes.
The number of extra dimensions increases from 0 to 7 as the curves
moved from top to bottom at high $\omega$.
Black hole masses corresponding to the non-commutative black hole maximum
temperature have been used.
$M_D = \sqrt{\theta} = 1$ has been taken.
}
\end{figure*}

The differences in Figure~\ref{fig:cross0} at low frequency are
predominately due to differences in horizon area.
It is enlightening to effectively remove these by scaling the cross
sections by $4\pi r_\mathrm{h}^2$ as shown in Figure~\ref{fig:cross1}.
The cross sections are now in better agreement for
$\omega \lesssim 0.25$ but have almost constant residual
differences for $\omega \gtrsim 0.25$.
These differences are due to the universal nature of the cross
section -- to be discussed later. 

\begin{figure*}[htb]
\includegraphics[width=\linewidth]{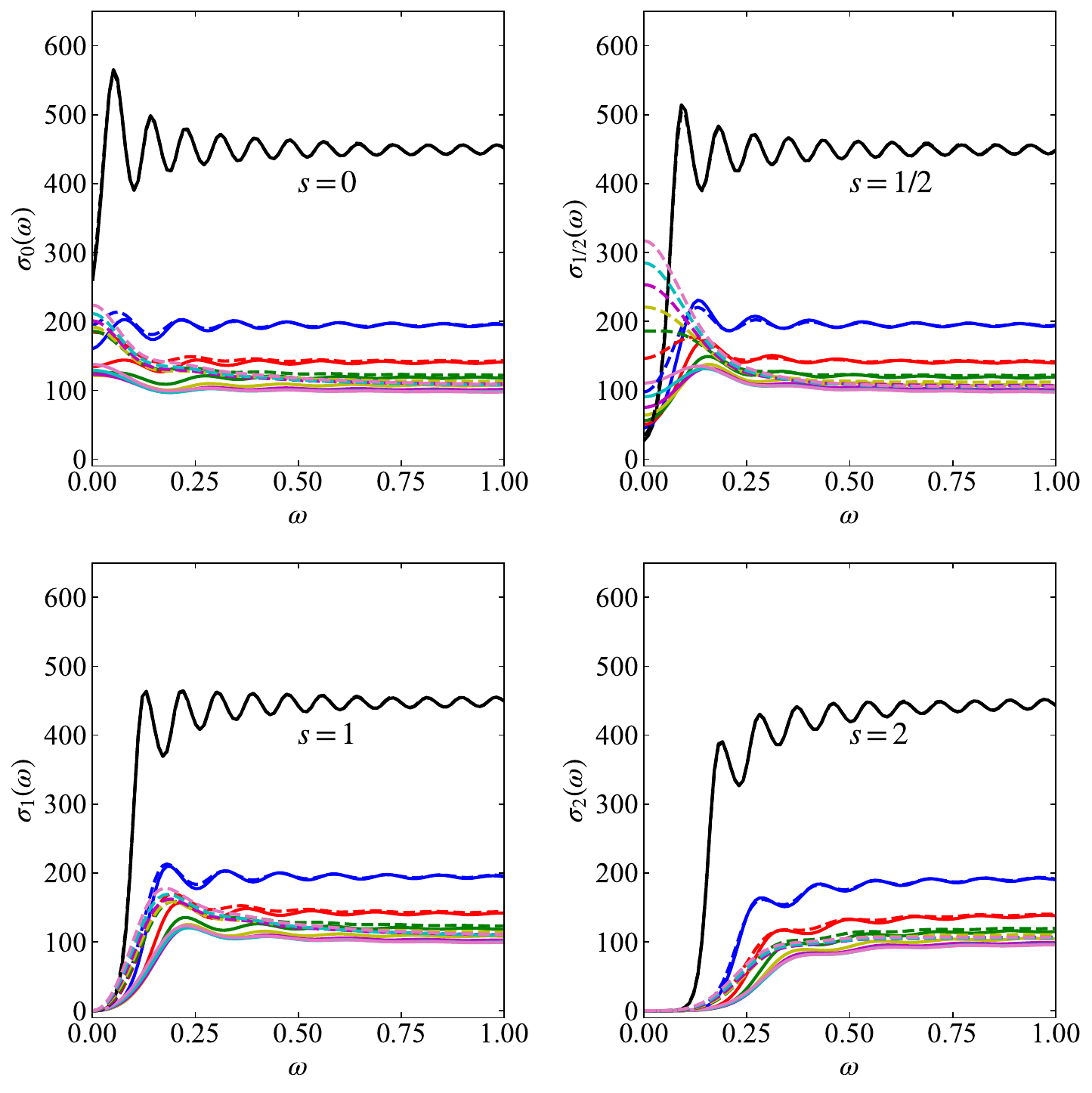}
\caption{\label{fig:cross2}%
Absorption cross sections versus frequency $\omega$ for $s = 0, 1/2,
1, 2$.   
The solid lines are for non-commutative black holes and dashed lines
for Schwarzschild-Tangherlini black holes.
The number of extra dimensions increases from 0 to 7 as the curves
moved from top to bottom at high $\omega$.
Black hole masses corresponding to the non-commutative black hole
minimum mass $+ 10M_D$ have been used.%
$M_D = \sqrt{\theta} = 1$ has been taken.
}
\end{figure*}

At the mass giving maximum NC black hole temperature, the horizon
radius of the NC and ST black holes are similar.
To examine larger differences due to the transmission
coefficients, we take masses near the minimum NC back hole horizon;
values from Table~\ref{tab:app} $+ 10M_D$ (for numerical stability).
Figure~\ref{fig:cross2} shows significant differences for all but the
low $n$ cases.

For ST black holes, the absorption cross section results are the same
as Ref.~\cite{Harris:2003eg}.
Although the absorption cross section results for NC black holes and
$s=0$ agree qualitatively with Ref.~\cite{Nicolini:2011nz}, they are
quantitatively different~\footnote{%
We do not understand the normalization of
Figure~7 in Ref.~\cite{Nicolini:2011nz}.
Figure~3 Ref.~\cite{Nicolini:2011nz} shows values of
$r_\mathrm{h} \sim 4.7-5.7$ at the maximum temperature.
These values correspond to horizon areas of about $4\pi r_\mathrm{h}^2
\sim 180-400$ which are in contradiction to what is shown in Figure~7 
Ref.~\cite{Nicolini:2011nz}.% 
}. 

The transmission coefficients for spin 0 and 1/2 turn on immediately
for tiny frequencies, leading to finite absorption cross sections at zero
frequency.
Transmission coefficient for spin 1 and 2 are essential zero at
$\omega = 0$, leading to an absorption cross section of zero at zero
frequency. 
The usual oscillations are seen and the number of peaks correspond to
the number of $\ell$ modes.
The oscillation are more predominate at low $n$ where the transmission
coefficients rise the steepest.
After normalizing by the horizon radius, the absorption cross sections
for NC black holes are higher than the ST back hole.

For spin 0, the low-frequency limit should correspond to the area of
the black hole for both ST and NC black holes: $\sigma_0^{(0)}= 4\pi
r_\mathrm{h}^2$. 
Numerically, for $\omega = 0.001$, we obtain the black
hole area to better than 0.9\% for both ST and NC black holes for high
and low mass and for all number of dimensions and spins; except for
$n=0$ for which it agrees to 1.7\% for ST black holes of mass
$M_{\mathrm{e}T}$. 
For spin 1/2, the low-frequency limit is given by \cite{Kanti:2002ge}

\begin{equation}
\sigma_0^{(1/2)} = 2^\frac{n-3}{n+1} 4\pi r_\mathrm{S}^2\, ,
\end{equation}

\noindent
which we are able to reproduce numerically to better than 0.4\%,
except for the $n=0$ case in which we obtain 2\% agreement.
We also mention that we obtain zero absorption cross section at
$\omega = 0.001$ for spin 1 and 2 fields from NC and ST 
black holes to 0.01\%.

In the high-frequency limit, it has been shown that the absorption
cross section approaches a universal geometrical optics limit of
$\sigma_\infty = \pi b_c^2$, where $b_c = r_c / \sqrt{h(r_c)}$ and
$r_c$ is given by the solution to $r_c h^\prime(r_c) - 2h(r_c) =
0$ \cite{Decanini:2011xi}. 
Using the ST metric, one obtains

\begin{equation}
\sigma_\infty = \left( \frac{n+3}{2} \right)^\frac{2}{n+1} \frac{n+3}{n+1}
\pi r_\mathrm{S}^2\, , 
\end{equation}

\noindent
where $r_\mathrm{S}$ is the ST horizon radius.
This result was first obtained a long time ago~\cite{Emparan:2000rs}.
In the case of the NC metric, we have calculated $\sigma_\infty$
numerically.
For $\omega = 1$, we obtain the optical cross section for ST and NC
black holes to better than 3\% for high mass, all number dimensions,
and all spins.
For low mass, the NC accuracy remains but the ST $n=7$ and $s=1$ case
worsens by up to 5\%. 
Visually, we already approach the geometrical limit for
$\omega \gtrsim 0.25$. 
Reproducing these known analytical values is a good test of the
numerical validity of our calculations.

%%%%%%%%%%%%%%%%%%%%%%%%%%%%%%%%%%%%%%%%%%%%%%%%%%%%%%%%%%%%%%%%%%%%%%%%%%%%%%%%%%
\subsection{Particle spectra}

\begin{figure*}[htb]
\includegraphics[width=\linewidth]{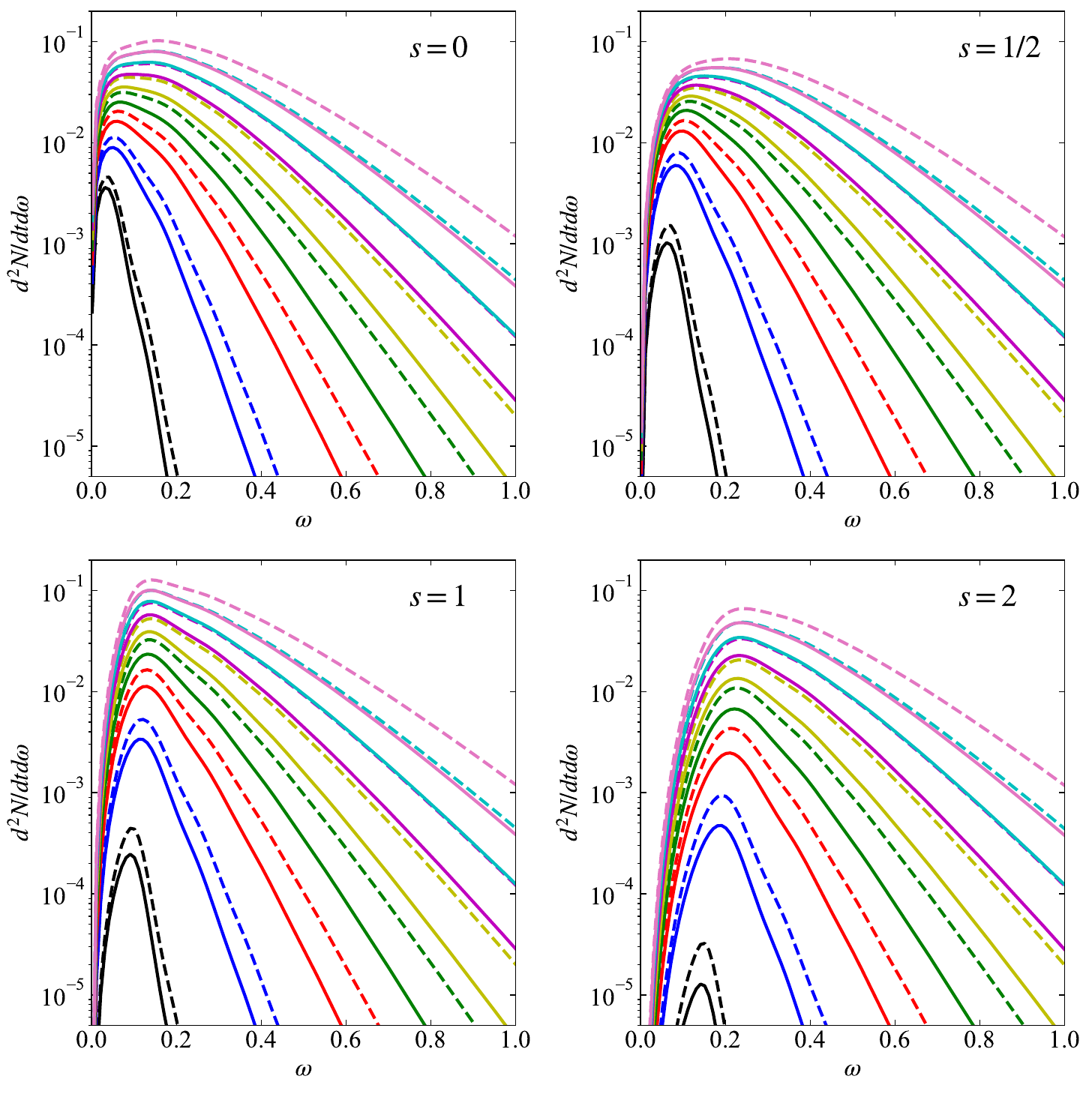}
\caption{\label{fig:NspectraeM}%
Particle spectra versus frequency $\omega$ for $s = 0, 1/2, 1, 2$.   
The solid lines are for non-commutative black holes and dashed lines
for Schwarzschild-Tangherlini black holes.
The number of extra dimensions increases from 0 to 7 as the curves
move from bottom to top.
Black hole masses corresponding to the non-commutative black hole maximum
temperature have been used.
$M_D = \sqrt{\theta} = 1$ has been taken.
}
\end{figure*}

The particle spectra have an additional dependence on temperature and
the dependence on frequency is only indirectly through the sum of
transmission coefficients and the statistical factor.
This time it is not possible to take the minimum NC black hole mass
(zero temperature) as the spectra will vanish due to the statistical factor.
An interesting choice is to take the NC black hole mass at its maximum
temperature.
For the ST black hole comparison, logical choices are to take the same mass
or the mass that gives the same temperature.
If the same temperature is taken, the statistical factor in the
particle spectra will be identical and the only difference will be the
transmission coefficient sum part of the formula.
First, we consider the case of equal mass which means the temperature
of the ST black hole will be hotter, and hence lead to significantly
more particle flux.  
Figure~\ref{fig:NspectraeM} shows
particle spectra versus frequency for $s = 0, 1/2, 1$, and 2.   
The solid lines are for NC black holes and dashed lines for ST black holes.
The number of extra dimensions increases from 0 to 7 as the curves
move from bottom to top.
Black hole masses $M_{\mathrm{e}M}$ corresponding to the NC black hole
maximum temperature have been used, as shown in
Table~\ref{tab:temperature}. 

\begin{figure*}[htb]
\includegraphics[width=\linewidth]{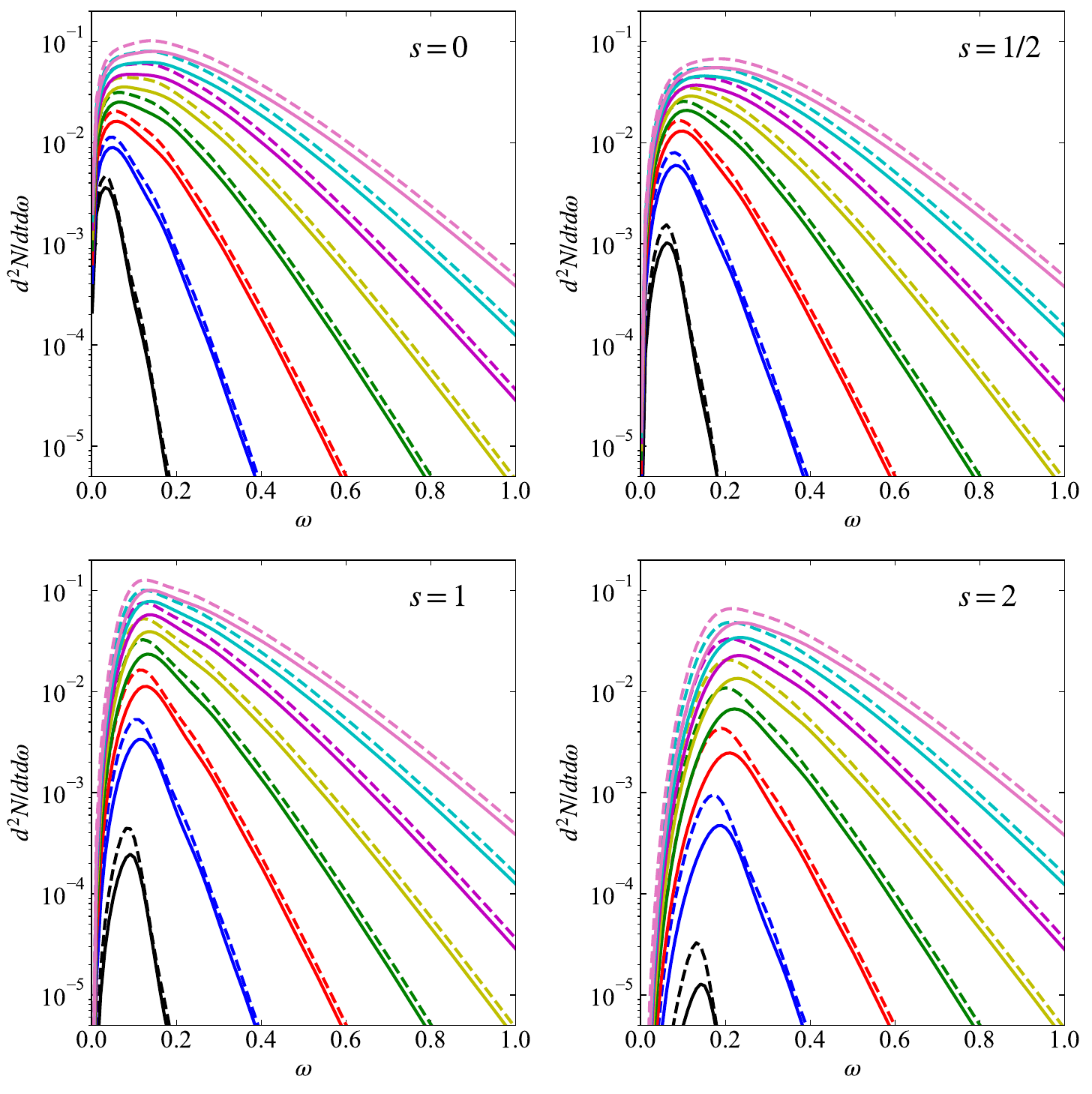}
\caption{\label{fig:NspectraeT}%
Particle spectra versus frequency $\omega$ for $s = 0, 1/2, 1, 2$.   
The solid lines are for non-commutative black holes and dashed lines
for Schwarzschild-Tangherlini black holes.
The number of extra dimensions increases from 0 to 7 as the curves
move from bottom to top.
Black hole masses corresponding the same temperature as the
non-commutative black hole maximum temperature have been used.
$M_D = \sqrt{\theta} = 1$ has been taken.
}
\end{figure*}

To remove the temperature dependence, different mass NC and ST black
holes are compared.
Figure~\ref{fig:NspectraeT} shows particle spectra versus frequency.
Black hole masses $M_{\mathrm{e}M}$ for NC black holes and
$M_{\mathrm{e}T}$ for ST black holes corresponding to the NC black
hole maximum temperature have been used, as shown in  
Table~\ref{tab:temperature}. 

%%%%%%%%%%%%%%%%%%%%%%%%%%%%%%%%%%%%%%%%%%%%%%%%%%%%%%%%%%%%%%%%%%%%%%%%%%%%%%%%%%
\subsection{Energy spectra}

\begin{figure*}[htb]
\includegraphics[width=\linewidth]{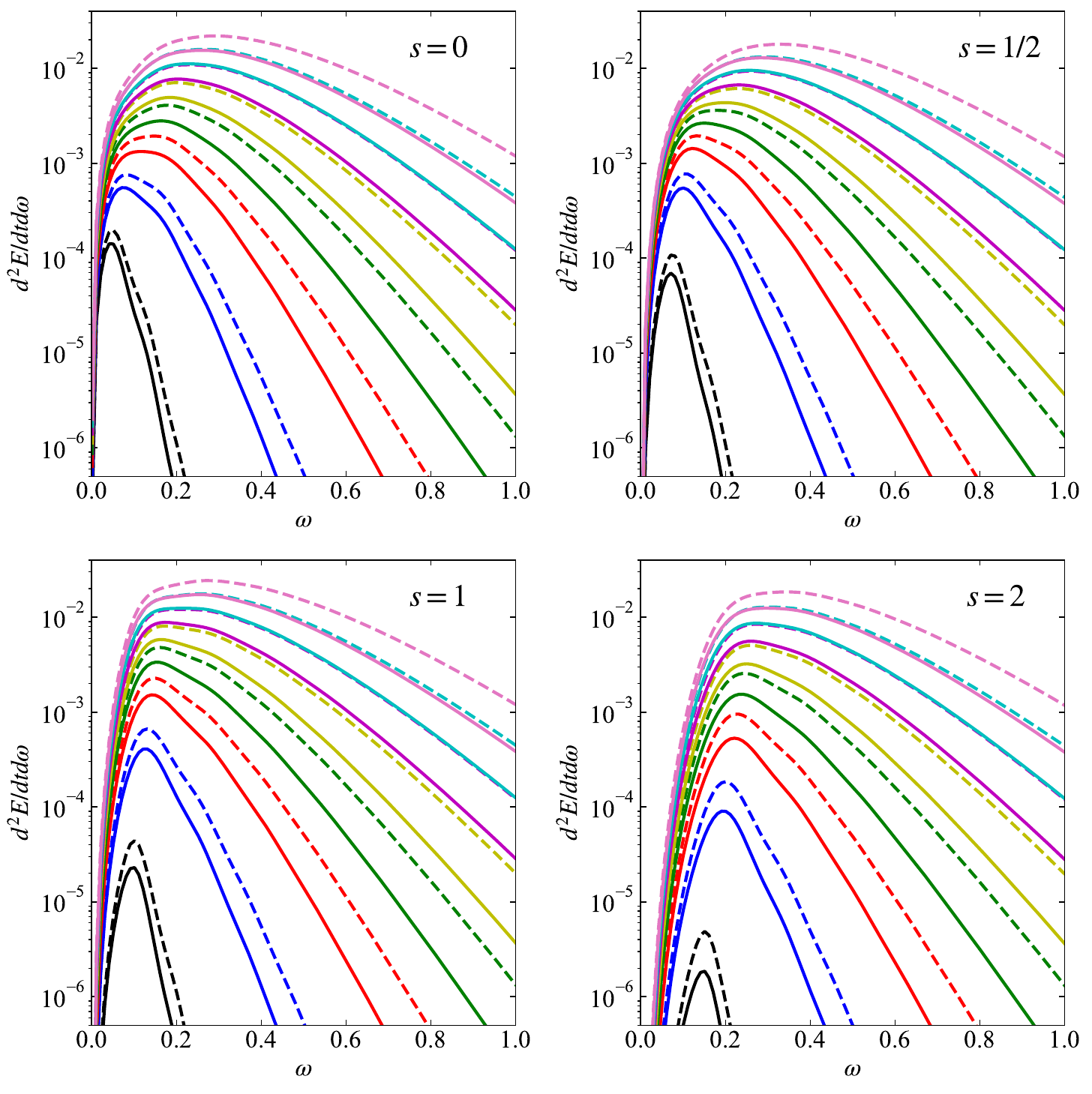}
\caption{\label{fig:EspectraeM}%
Energy spectra versus frequency $\omega$ for $s = 0, 1/2, 1, 2$.  
The solid lines are for non-commutative black holes and dashed lines
for Schwarzschild-Tangherlini black holes.
The number of extra dimensions increases from 0 to 7 as the curves
move from bottom to top.
Black hole masses corresponding to the non-commutative black hole maximum
temperature have been used.
$M_D = \sqrt{\theta} = 1$ has been taken.
}
\end{figure*}

The energy spectra are similar to the particle spectra but include a
multiplicative frequency factor.
Figure~\ref{fig:EspectraeM} shows energy spectra versus frequency
for $s = 0, 1/2, 1$ and 2.   
The solid lines are for NC black holes and dashed lines for ST black holes.
The number of extra dimensions increases from 0 to 7 as the curves
move from bottom to top.
Black hole masses $M_{\mathrm{e}M}$ corresponding to the NC black hole
maximum temperature have been used, as shown in
Table~\ref{tab:temperature}. 

\begin{figure*}[htb]
\includegraphics[width=\linewidth]{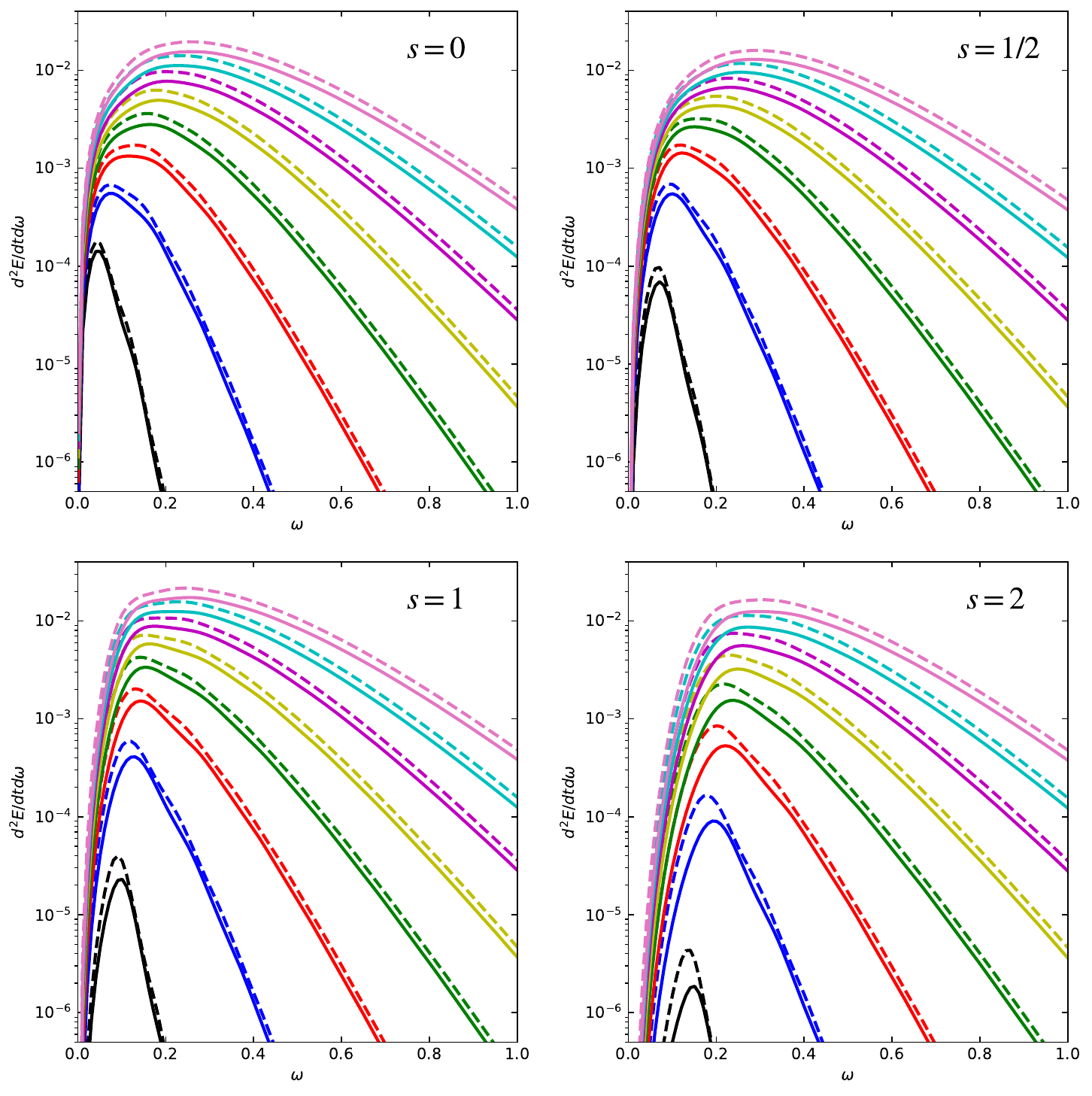}
\caption{\label{fig:EspectraeT}%
Energy spectra versus frequency $\omega$ for $s = 0, 1/2, 1, 2$.  
The solid lines are for non-commutative black holes and dashed lines
for Schwarzschild-Tangherlini black holes.
The number of extra dimensions increases from 0 to 7 as the curves
move from bottom to top.
Black hole masses corresponding the same temperature as the
non-commutative black hole maximum temperature have been used.
$M_D = \sqrt{\theta} = 1$ has been taken.
}
\end{figure*}

To remove the temperature dependence, different mass NC and ST black
holes are compared.
Figure~\ref{fig:EspectraeT} shows energy spectra versus frequency.
Black hole masses $M_{\mathrm{e}M}$ for NC black holes and
$M_{\mathrm{e}T}$ for ST black holes corresponding to the NC black
hole maximum temperature have been used, as shown in  
Table~\ref{tab:temperature}. 

%%%%%%%%%%%%%%%%%%%%%%%%%%%%%%%%%%%%%%%%%%%%%%%%%%%%%%%%%%%%%%%%%%%%%%%%%%%%%%%%%%
\subsection{Particle flux and total power}

To make the comparison quantitative, we integrate the particle spectra
and energy spectra over frequency out to $\omega = 1$ to
obtain the particle flux and power, respectively.
Table~\ref{tab:FcompeM} and Table~\ref{tab:FcompeT} show the NC to ST
particle flux ratios for the cases of equal mass and equal temperature,
respectively.
We observe the ratio of spin 0 and 1 fields are not very sensitive to
number of extra dimensions for $n>0$.
The biggest change in particle flux ratio with number of dimensions is
for spin 2.

\begin{table}[htb]
\caption{\label{tab:FcompeM}%
Ratio of particle flux from non-commutative to
Schwarzschild-Tangherlini black holes versus spin $s$ and number of
extra dimensions $n$ at the mass of the non-commutative black hole
of maximum temperature, $M_{\mathrm{e}M}$. 
}
\begin{indented}
\item[]\begin{tabular}{ccccccccc}
\br
& \multicolumn{8}{c}{$n$}\\
$s$ & 0 & 1 & 2 & 3 & 4 & 5 & 6 & 7\\
\mr
0   & 0.73 & 0.71 & 0.70 & 0.70 & 0.70 & 0.70 & 0.70 & 0.71\\
1/2 & 0.65 & 0.69 & 0.70 & 0.71 & 0.71 & 0.72 & 0.72 & 0.72\\
1   & 0.54 & 0.60 & 0.63 & 0.65 & 0.67 & 0.68 & 0.69 & 0.69\\
2   & 0.40 & 0.49 & 0.53 & 0.57 & 0.59 & 0.61 & 0.63 & 0.64\\
\br
\end{tabular}
\end{indented}
\end{table}

\begin{table}[htb]
\caption{\label{tab:FcompeT}%
Ratio of particle flux from non-commutative to
Schwarzschild-Tangherlini black holes versus spin $s$ and number of
extra dimensions $n$ at masses corresponding to the non-commutative
black hole maximum temperature: $M_{\mathrm{e}M}$ for non-commutative and
$M_{\mathrm{e}T}$ for Schwarzschild-Tangherlini black holes. 
}
\begin{indented}
\item[]\begin{tabular}{ccccccccc}
\br
& \multicolumn{8}{c}{$n$}\\
$s$ & 0 & 1 & 2 & 3 & 4 & 5 & 6 & 7\\
\mr
0   & 0.80 & 0.79 & 0.79 & 0.79 & 0.79 & 0.79 & 0.79 & 0.79\\
1/2 & 0.71 & 0.77 & 0.79 & 0.80 & 0.80 & 0.81 & 0.81 & 0.81\\
1   & 0.59 & 0.67 & 0.71 & 0.74 & 0.75 & 0.76 & 0.77 & 0.78\\
2   & 0.44 & 0.54 & 0.60 & 0.64 & 0.67 & 0.69 & 0.71 & 0.72\\
\br
\end{tabular}
\end{indented}
\end{table}

Table~\ref{tab:PcompeM} and Table~\ref{tab:PcompeT} show the NC to ST
power ratios for the cases of equal mass and equal temperature,
respectively.
The same observations can be made as for the particle fluxes.

\begin{table}[htb]
\caption{\label{tab:PcompeM}%
Ratio of power emitted from non-commutative to
Schwarzschild-Tangherlini black holes versus spin $s$ and number of
extra dimensions $n$ at the mass of the non-commutative black hole
of maximum temperature, $M_{\mathrm{e}M}$. 
}
\begin{indented}
\item[]\begin{tabular}{ccccccccc}
\br
& \multicolumn{8}{c}{$n$}\\
$s$ & 0 & 1 & 2 & 3 & 4 & 5 & 6 & 7\\
\mr
0   & 0.67 & 0.64 & 0.62 & 0.62 & 0.62 & 0.62 & 0.63 & 0.63\\
1/2 & 0.61 & 0.63 & 0.63 & 0.63 & 0.63 & 0.63 & 0.64 & 0.64\\
1   & 0.51 & 0.57 & 0.59 & 0.60 & 0.61 & 0.62 & 0.63 & 0.63\\
2   & 0.39 & 0.46 & 0.50 & 0.53 & 0.55 & 0.57 & 0.59 & 0.60\\
\br
\end{tabular}
\end{indented}
\end{table}

\begin{table}[htb]
\caption{\label{tab:PcompeT}%
Ratio of power emitted from non-commutative to
Schwarzschild-Tangherlini black holes versus spin $s$ and number of
extra dimensions $n$ at masses corresponding to the non-commutative
black hole maximum temperature: $M_{\mathrm{e}M}$ for non-commutative and
$M_{\mathrm{e}T}$ for Schwarzschild-Tangherlini black holes. 
}
\begin{indented}
\item[]\begin{tabular}{ccccccccc}
\br
& \multicolumn{8}{c}{$n$}\\
$s$ & 0 & 1 & 2 & 3 & 4 & 5 & 6 & 7\\
\mr
0   & 0.82 & 0.80 & 0.79 & 0.79 & 0.79 & 0.79 & 0.79 & 0.80\\
1/2 & 0.74 & 0.79 & 0.80 & 0.80 & 0.80 & 0.80 & 0.81 & 0.81\\
1   & 0.63 & 0.71 & 0.75 & 0.77 & 0.78 & 0.79 & 0.79 & 0.80\\
2   & 0.47 & 0.58 & 0.64 & 0.68 & 0.70 & 0.72 & 0.74 & 0.75\\
\br
\end{tabular}
\end{indented}
\end{table}

Concentrating on NC geometry inspired black holes, we calculate the
particle flux and total power for each number of extra dimensions and
compare it to the $n=0$ case shown in Table~\ref{tab:flux} and
Table~\ref{tab:power}, respectively.
Direct comparison with Ref.~\cite{Harris:2003eg} of the results for ST
black holes (not shown) is not possible since the black hole mass used
is not stated.
The results are however, similar.

\begin{table}[htb]
\caption{\label{tab:flux}%
Particle flux ratios for different number of extra dimensions $n$
relative to $n=0$ versus spin $s$ for non-commutative black holes with
the maximum temperature. 
}
\begin{indented}
\item[]\begin{tabular}{ccccccccc}
\br
& \multicolumn{8}{c}{$n$}\\
$s$ & 0 & 1 & 2 & 3 & 4 & 5 & 6 & 7\\
\mr
0   & 1 &  5 &  12 &   25 &   44 &   69 &   101 &   141\\
1/2 & 1 & 10 &  29 &   61 &  104 &  160 &   230 &   312\\
1   & 1 & 22 &  93 &  235 &  459 &  772 &  1179 &  1685\\
2   & 1 & 59 & 406 & 1354 & 3189 & 6149 & 10412 & 16131\\
\br
\end{tabular}
\end{indented}
\end{table}

\begin{table}[htb]
\caption{\label{tab:power}%
Power emission ratios for different number of extra dimensions $n$
relative to $n=0$ versus spin $s$ for non-commutative black holes with
the maximum temperature. 
}
\begin{indented}
\item[]\begin{tabular}{ccccccccc}
\br
& \multicolumn{8}{c}{$n$}\\
$s$ & 0 & 1 & 2 & 3 & 4 & 5 & 6 & 7\\
\mr
0   & 1 &  9 &  32 &   84 &  174 &   316 &   519 &   794\\
1/2 & 1 & 15 &  60 &  153 &  311 &   551 &   887 &  1334\\
1   & 1 & 30 & 160 &  469 & 1031 &  1916 &  3187 &  4899\\
2   & 1 & 82 & 668 & 2512 & 6494 & 13504 & 24379 & 39872\\
\br
\end{tabular}
\end{indented}
\end{table}

Shown in Table~\ref{tab:Fspin} and Table~\ref{tab:Pspin} are the case
of particle flux and power for each spin compared to the spin-0 case.
Direct comparison with Ref.~\cite{Harris:2003eg} of the results for ST
black holes (not shown) is not possible since the black hole mass used
is not stated.
However, the results are the same as Ref.~\cite{Harris:2003eg} for most
cases, except for a difference of 1\% for some $n=7$ spins.

\begin{table}[htb]
\caption{\label{tab:Fspin}%
Particle flux ratios for different spin $s$ relative to $s=0$ versus
number of extra dimensions $n$ for non-commutative black holes with
the maximum temperature. 
}
\begin{indented}
\item[]\begin{tabular}{ccccc}
\br
$n$ & $s=0$ & $s=1/2$ & $s=1$ & $s=2$\\
\mr
0 & 1 & 0.33 & 0.08 & 0.005\\
1 & 1 & 0.68 & 0.38 & 0.06\\
2 & 1 & 0.78 & 0.62 & 0.15\\
3 & 1 & 0.79 & 0.77 & 0.25\\
4 & 1 & 0.78 & 0.87 & 0.33\\
5 & 1 & 0.76 & 0.93 & 0.41\\
6 & 1 & 0.74 & 0.97 & 0.47\\
7 & 1 & 0.73 & 0.99 & 0.53\\
\br
\end{tabular}
\end{indented}
\end{table}

\begin{table}[htb]
\caption{\label{tab:Pspin}%
Power emission ratios for different spin $s$ relative to $s=0$ versus
number of extra dimensions $n$ for non-commutative black holes with
the maximum temperature. 
}
\begin{indented}
\item[]\begin{tabular}{ccccc}
\br
$n$ & $s=0$ & $s=1/2$ & $s=1$ & $s=2$\\
\mr
0 & 1 & 0.50 & 0.17 & 0.01\\
1 & 1 & 0.86 & 0.61 & 0.14\\
2 & 1 & 0.92 & 0.86 & 0.30\\
3 & 1 & 0.91 & 0.97 & 0.44\\
4 & 1 & 0.89 & 1.03 & 0.55\\
5 & 1 & 0.87 & 1.05 & 0.63\\
6 & 1 & 0.85 & 1.06 & 0.69\\
7 & 1 & 0.84 & 1.07 & 0.74\\
\br
\end{tabular}
\end{indented}
\end{table}
%%%%%%%%%%%%%%%%%%%%%%%%%%%%%%%%%%%%%%%%%%%%%%%%%%%%%%%%%%%%%%%%%%%%%%%%%%%%%%%%%%
\section{Discussion}

Transmission coefficients for spin 0, 1/2, 1, and 2 fields from NC
geometry inspired black holes of extra dimension from 0 to 7 have been
calculated.
The NC black hole transmission coefficients are similar to the ST
black hole transmission coefficients when their horizon radius are
similar.
However, there are major differences when the black hole masses are
similar but the horizon radius are significantly different.
The major difference in transmission coefficients is that the NC black
hole transmission coefficients turn-on at slightly higher frequency.
The differences between NC black hole and ST black hole transmission
coefficients are about the same for all spins.

The absorption cross section of different spin fields from NC
geometry inspired black holes of different number extra dimensions
have been calculated. 
Significant differences in NC black hole and ST black hole absorption
cross sections occur at low frequencies while the cross sections at
high frequencies approach the geometrical optics limits.
For masses near the minimum NC black hole mass, the differences are
more apparent, particularly for higher dimensions.

The particle and energy spectra on the brane of different spin fields
from NC geometry inspired black holes of different number of extra
dimensions have been calculated.
For equal masses, the NC black hole spectra are significantly lower
than for ST black holes, mainly due to the lower temperature.
However, at equal temperature the NC black hole spectra are still
significantly lower than for ST black hole spectra.

When integrating the particle spectra over frequency the particle flux
from NC black holes is significantly less than that from ST black
holes.
For spin-0 fields, the reduction can be from 0.70-0.80 depending on if
the black holes have equal mass (lower number) or equal temperature
(higher number).
The dependence on the number of dimensions is small.
For spin-1/2 fields, the ratio is from 0.65-0.81, also not too
dependent on number of extra dimensions.
For spin-1 fields, the ratio is 0.54-0.78.
But for spin-2 fields, the reduction is most significant from
0.40-0.72, increasing with increasing number of dimensions.
The general trends in the power are similar to the trends in the
particle flux.

Considering the NC geometry inspired black hole emission on its own,
it is common to compare the fluxes to the spin-0 case or the $n=0$
case.
Increases in particle flux and power relative to the $n=0$ case are
observed for increasing number of dimensions.
The increase is most prominent for spin 2 and smallest for spin 0.
The particle flux and power of spin-1/2 fields relative to spin-0
fields is less, with $n=0$ being the lowest and $n=3$ the highest for
the particle flux and $n=2$ for power.
The particle flux of spin-1 fields relative to spin-0 fields is less
and decreases significantly with decreasing number of dimensions.
The power of spin-1 fields relative to spin-0 fields is less or
greater, depending on the number of dimensions; being significantly
less for $n=0$ and slightly greater for $n=7$.
The spin-2 fields particle fluxes and power are always significantly
less than for spin-0 fields, being 1\% for $n=0$.

We have presented greybody factors, absorption cross sections, and
particle and energy spectra for all spin fields from
higher-dimensional non-commutative geometry inspired black holes for
the first time.
The calculations are numerical and thus valid over the entire
frequency range.
The emission of higher spin fields, particularly graviton emission,
could be useful for relating possible future observations of
high-temperature black hole radiation to theory.
The reduction in emission due to the greybody factors, not
temperature, that we observe are hopefully model independent.
This work represents another step towards possibly elucidating some
aspects of quantum gravity. 

%%%%%%%%%%%%%%%%%%%%%%%%%%%%%%%%%%%%%%%%%%%%%%%%%%%%%%%%%%%%%%%%%%%%%%%
\section*{Acknowledgments}
We acknowledge the support of the Natural Sciences and Engineering
Research Council of Canada (NSERC). 
Nous remercions le Conseil de recherches en sciences naturelles et en
g{\'e}nie du Canada (CRSNG) de son soutien. 

%%%%%%%%%%%%%%%%%%%%%%%%%%%%%%%%%%%%%%%%%%%%%%%%%%%%%%%%%%%%%%%%%%%%%%%
\appendix
\section{Experimental constraints\label{sec:LHC}}

The experimental lower bounds on $M_D$ and the maximum energy of the
LHC will restrict the values of $\sqrt{\theta}$ that can be probed by
experiments at the LHC. 
We do not expect black holes to form for masses much less than $M_D$.
This give a lower bound on $M$.
We will consider only the hard limits on the Planck scale set by
accelerator experiments~\cite{D0:2008ayi,ATLAS:2021kxv}:
$M_D > 11.2$~TeV for $n = 2$,
$M_D > 8.5$~TeV for $n = 3$,
$M_D > 7.1$~TeV for $n = 4$,
$M_D > 6.4$~TeV for $n = 5$,
$M_D > 5.9$~TeV for $n = 6$, and
$M_D > 0.8$~TeV for $n = 7$.
The maximum mass of the black hole is likely to be limited by the
statistics of the maximum parton energies in a proton-proton
interaction but in no case can it be larger than the proton-proton
center-of-mass energy.
Thus, we will only be interested in the case where the minimum black
hole mass is below the LHC current maximum energy of 13.6~TeV and
above the experimental lower bound on the Planck scale.

\begin{table}[htb]
\caption{\label{tab:app}%
Values of minimum horizon radius $(r_\mathrm{h})_\mathrm{min}$
in units of $\sqrt{\theta}$ and minimum mass $M_\mathrm{min}$ in units
of $M_D(\sqrt{\theta}M_D)^{n+1}$.
The last two columns show the range of $\sqrt{\theta}$ in units of
$1/M_D$ that can be probed at the Large Hadron Collider.
}
\begin{indented}
\item[]\begin{tabular}{ccccc}
\br
$n$ & $(r_\mathrm{h})_\mathrm{min}/\sqrt{\theta}$ &
$\frac{M_\mathrm{min}/M_D}{(\sqrt{\theta}M_D)^{n+1}}$ &
  $\sqrt{\theta_\mathrm{min}}M_D$ & $\sqrt{\theta_\mathrm{max}}M_D$\\
\mr
0 & 3.02 & 47.9 & &\\
1 & 2.68 & 63.2 & &\\
2 & 2.51 & 65.2 & 0.248 & 0.265\\
3 & 2.41 & 58.8 & 0.361 & 0.406\\
4 & 2.34 & 48.6 & 0.460 & 0.524\\
5 & 2.29 & 37.9 & 0.546 & 0.619\\
6 & 2.26 & 28.2 & 0.621 & 0.699\\
7 & 2.23 & 20.3 & 0.686 & 0.978\\
\br
\end{tabular}
\end{indented}
\end{table}

We obtain a valid range of $\sqrt{\theta} M_D$ for each number of
extra dimensions by restricting the minimum black hole mass at the LHC
to be in the range $1 < M_\mathrm{min}/M_D < 13.6$~TeV/$M_D$, as
discussed above.
The results are given in Table~\ref{tab:app}.
We see that $\sqrt{\theta}$ is very restricted and there is no single
value of $\sqrt{\theta}$ that lies in the allowed range for all number
of extra dimensions.

To study the phenomenology of NC inspired black holes at the LHC
experiments one can take $M_D$ above the experimental limits and the
following values 
$\sqrt{\theta} = 0.3$ for $n=2$,
$\sqrt{\theta} = 0.4$ for $n=3$,
$\sqrt{\theta} = 0.5$ for $n=4$,
$\sqrt{\theta} = 0.6$ for $n=5$,
$\sqrt{\theta} = 0.7$ for $n=6$, and
$\sqrt{\theta} = 0.8$ for $n=7$.

%%%%%%%%%%%%%%%%%%%%%%%%%%%%%%%%%%%%%%%%%%%%%%%%%%%%%%%%%%%%%%%%%%%%%%%
\section*{References}
\bibliographystyle{unsrt}
\bibliography{gingrich}

%%%%%%%%%%%%%%%%%%%%%%%%%%%%%%%%%%%%%%%%%%%%%%%%%%%%%%%%%%%%%%%%%%%%%%%
\end{document}